\documentclass[12pt,english]{article}

\usepackage{amsmath}
\usepackage{amssymb}
\usepackage{graphicx}
\usepackage{cite}

\makeatletter

 \@addtoreset{equation}{section}

\allowdisplaybreaks[1]

\newif\ifContLineOne
\newif\ifContLineTwo
\newif\ifContLineThree

\def\conC#1{\vbox{\ialign{##\crcr
  \ifContLineThree\hrulefill\else\vphantom{\hrulefill}\fi\crcr
  \noalign{\kern3.2pt\nointerlineskip}
  \ifContLineTwo\hrulefill\else\vphantom{\hrulefill}\fi\crcr
  \noalign{\kern3.2pt\nointerlineskip}
  \ifContLineOne\hrulefill\else\vphantom{\hrulefill}\fi\crcr
  \noalign{\nointerlineskip}
  $\hfil\textstyle{\vbox to 14pt{}#1}\hfil$\crcr}}}

\def\DrawLeg#1#2{
  \kern-.2pt              
  \dimen2 =#1             
  \advance\dimen2 by 2pt  
  \dimen3 = 10.6pt        
  \dimen4 =3.6pt          
  \advance\dimen3 by -\dimen2 
  \multiply\dimen4 by #2
  \advance\dimen3 by \dimen4
  \raise\dimen2 \hbox{\vrule height\dimen3 width .4pt} 
  \kern-.2pt}             

\def\begC#1#2{\setbox0 =\hbox{$\textstyle{#2}$}
  \dimen0=.5\wd0 \dimen1=\ht0
  \conC{\hskip\dimen0}
  \count255=#1
  \ifnum\count255 =1 \ContLineOnetrue\else
  \ifnum\count255 =2 \ContLineTwotrue\else
  \ifnum\count255 =3 \ContLineThreetrue\fi\fi\fi
  \DrawLeg{\dimen1}{\count255}
  \conC{\hskip\dimen0}
  \kern-\dimen0\kern-\dimen0 \box0}

\def\endC#1#2{\setbox0 =\hbox{$\textstyle{#2}$}
  \dimen0=.5\wd0 \dimen1=\ht0
  \conC{\hskip\dimen0}
  \count255=#1
  \ifnum\count255 =1 \ContLineOnefalse\else
  \ifnum\count255 =2 \ContLineTwofalse\else
  \ifnum\count255 =3 \ContLineThreefalse\fi\fi\fi
  \DrawLeg{\dimen1}{\count255}
  \conC{\hskip\dimen0}
  \kern-\dimen0\kern-\dimen0 \box0}

\makeatother

\usepackage{babel}

\begin{document}

\begin{titlepage}

\global\long\def\thefootnote{\fnsymbol{footnote}}

\begin{flushright}
UTHEP-710
\end{flushright}

\bigskip{}

\begin{center}
\textbf{\Large  
Multiloop Amplitudes of Light-cone Gauge Superstring Field Theory:
Odd Spin Structure Contributions
 }
\par\end{center}

\bigskip{}

\begin{center}
{\large 
Nobuyuki Ishibashi$^{a}$\footnote{e-mail: ishibash@het.ph.tsukuba.ac.jp}
and Koichi Murakami$^{b}$\footnote{e-mail: koichi@kushiro-ct.ac.jp} 
}
\end{center}

\begin{center}
$^{a}$\textit{ Tomonaga Center for the History of the Universe, }\\
\textit{University of Tsukuba}\\
\textit{Tsukuba, Ibaraki 305-8571, JAPAN}\\

\par\end{center}

\begin{center}
$^{b}$\textit{National Institute of Technology, Kushiro College,}\\
\textit{ Otanoshike-Nishi 2-32-1, Kushiro, Hokkaido 084-0916, JAPAN} 
\par\end{center}

\bigskip{}

\bigskip{}

\bigskip{}

\begin{abstract}
We study the odd spin structure contributions to the multiloop amplitudes
of light-cone gauge superstring field theory. We show that they coincide
with the amplitudes in the conformal gauge with two of the vertex
operators chosen to be in the pictures different from the standard
choice, namely $\left(-1,-1\right)$ picture in the type II case and
$-1$ picture in the heterotic case. We also show that the contact term
divergences can be regularized in the same way as in the amplitudes
for the even structures and we get the amplitudes which coincide with
those obtained from the first-quantized approach. 
\end{abstract}
\global\long\def\thefootnote{\arabic{footnote}}

\end{titlepage}

\pagebreak{}


\section{Introduction}

String field theory is expected to provide a nonperturbative formulation
of string theory. It is a second-quantized string theory from which
one can calculate Feynman amplitudes which agree with those of the
first-quantized theory. For bosonic strings, there are several proposals
of such string field theories. For superstrings, because of the problems
with the method to calculate multiloop amplitudes using the picture
changing operators, the construction of a string field theory has
been a difficult problem. Recently, Sen has constructed a gauge invariant
formulation of the string field theory for closed superstrings \cite{Sen2016c,Sen2015b,Sen2015j,Lacroix2017b,Sen2017},
based on the formulation \cite{Zwiebach1993} of closed string field
theory for bosonic strings with a nonpolynomial action. 

Light-cone gauge closed superstring field theory is a string field
theory for superstrings which involves only three-string interaction
terms. It can be proved formally that the Feynman amplitudes of the
string field theory coincide with those of the first-quantized theory
\cite{Aoki:1990yn}. The proof was formal because there appear unphysical
divergences which are called the contact term divergences 
\cite{Greensite:1986gv,Greensite:1987sm,Greensite:1987hm,Green:1987qu,Wendt:1987zh}. In a previous
paper \cite{Ishibashi2017b}, we have shown that these divergences
can be dealt with by dimensional regularization. In the case of type
II superstrings, for example, one formulates a light-cone gauge superstring
field theory in noncritical dimensions or the one whose worldsheet
theory for transverse variables is a superconformal field theory with
central charge $c\ne12$ \cite{Ishibashi2017d}. Although Lorentz
invariance is broken by doing so, it does not cause so much trouble
because the light-cone gauge theory is a completely gauge-fixed theory.
In \cite{Ishibashi2017b}, we have shown that the multiloop amplitudes
corresponding to the Riemann surfaces with even spin structure involving
external lines in the (NS,NS) sector can be calculated using the dimensional
regularization and the results coincide with those of the first-quantized
approach. 

What we would like to do in this paper is to generalize these results
to the case of the surfaces with odd spin structure. On the Riemann
surfaces with odd spin structure, there exist zero modes of the fermionic
variables on the worldsheet which make the manipulations of the amplitudes
complicated. We will show that it is possible to deal with these zero
modes and prove that the amplitudes are equal to those of the first-quantized
method, when all the external lines are in the (NS,NS) sector, in the case of type II superstrings. It
is straightforward to obtain similar results for heterotic strings.

The organization of this paper is as follows. In section \ref{sec:Light-cone-gauge-superstring},
we review the results in \cite{Ishibashi2017b} and the problems with
the odd spin structures. In section \ref{sec:Odd-spin-structure},
we deal with the amplitudes for the odd spin structure and show that
these also coincide with those from the first-quantized approach.
Section \ref{sec:Conclusions-and-discussions} is devoted to discussions.
In the appendices, we present details of the manipulations given in
the main text. 

\section{Light-cone gauge superstring field theory\label{sec:Light-cone-gauge-superstring}}

In this section, we review the known results for the multiloop amplitudes
of light-cone gauge superstring field theory and the problems with
the odd spin structures. 

\subsection{Light-cone gauge superstring field theory\label{subsec:Light-cone-gauge-superstring}}

In the light-cone gauge string field theory, the string field 
\begin{equation}
\left|\Phi\left(t,\alpha\right)\right\rangle 
\end{equation}
is taken to be an element of the Hilbert space $\mathcal{H}$ of the
transverse variables on the worldsheet and a function of 
\begin{eqnarray}
t & = & x^{+}\,,\nonumber \\
\alpha & = & 2p^{+}\,.
\end{eqnarray}
In this paper, we consider the string field theory for type II superstrings
in $10$ dimensional flat spacetime as an example. $\left|\Phi(t,\alpha)\right\rangle $
should be GSO even and satisfy the level-matching condition 
\begin{equation}
(L_{0}-\bar{L}_{0})\left|\Phi\left(t,\alpha\right)\right\rangle =0\,,\label{eq:levelmatching}
\end{equation}
where $L_{0},\bar{L}_{0}$ are the zero modes of the Virasoro generators
of the worldsheet theory.

The action of the string field theory is given by 
\begin{eqnarray}
S & = & \int dt\left[\frac{1}{2}\sum_{\mathrm{B}}\int_{-\infty}^{\infty}\frac{\alpha d\alpha}{4\pi}\left\langle \Phi_{\mathrm{B}}\left(-\alpha\right)\right|(i\partial_{t}-\frac{L_{0}+\bar{L}_{0}-1}{\alpha})\left|\Phi_{\mathrm{B}}\left(\alpha\right)\right\rangle \right.\nonumber \\
 &  & \ +\frac{1}{2}\sum_{\mathrm{F}}\int_{-\infty}^{\infty}\frac{d\alpha}{4\pi}\left\langle \Phi_{\mathrm{F}}\left(-\alpha\right)\right|(i\partial_{t}-\frac{L_{0}+\bar{L}_{0}-1}{\alpha})\left|\Phi_{\mathrm{F}}\left(\alpha\right)\right\rangle \nonumber \\
 &  & \ -\frac{g_{s}}{6}\sum_{\mathrm{B}_{1},\mathrm{B}_{2},\mathrm{B}_{3}}\int\prod_{r=1}^{3}\left(\frac{\alpha_{r}d\alpha_{r}}{4\pi}\right)\delta\left(\sum_{r=1}^{3}\alpha_{r}\right)\left\langle V_{3}\left|\Phi_{\mathrm{B}_{1}}(\alpha_{1})\right.\right\rangle \left|\Phi_{\mathrm{B}_{2}}(\alpha_{2})\right\rangle \left|\Phi_{\mathrm{B}_{3}}(\alpha_{3})\right\rangle \nonumber \\
 &  & \ \left.-\frac{g_{s}}{2}\sum_{\mathrm{B}_{1},\mathrm{F}_{2},\mathrm{F}_{3}}\int\prod_{r=1}^{3}\left(\frac{\alpha_{r}d\alpha_{r}}{4\pi}\right)\delta\left(\sum_{r=1}^{3}\alpha_{r}\right)\left\langle V_{3}\left|\Phi_{\mathrm{B}_{1}}(\alpha_{1})\right.\right\rangle \alpha_{2}^{-\frac{1}{2}}\left|\Phi_{\mathrm{F}_{2}}(\alpha_{2})\right\rangle \alpha_{3}^{-\frac{1}{2}}\left|\Phi_{\mathrm{F}_{3}}(\alpha_{3})\right\rangle \right],\nonumber \\
 &  & \ \label{eq:superHamiltonian}
\end{eqnarray}
which consists of the kinetic terms and the three-string interaction
terms. $\sum_{\mathrm{B}}$ and $\sum_{\mathrm{F}}$ denote the sums
over bosonic and fermionic string fields respectively. The three-string
vertices $\left\langle V_{3}\right|$ are elements of $\mathcal{H^{\ast}}\otimes\mathcal{H^{\ast}}\otimes\mathcal{H^{\ast}}$
whose definition can be found in \cite{Baba:2009kr,Ishibashi:2010nq,Ishibashi2017b}.
Here $\mathcal{H}^{\ast}$denotes the dual space of $\mathcal{H}$. 

\begin{figure}
\begin{centering}
\includegraphics[scale=0.5]{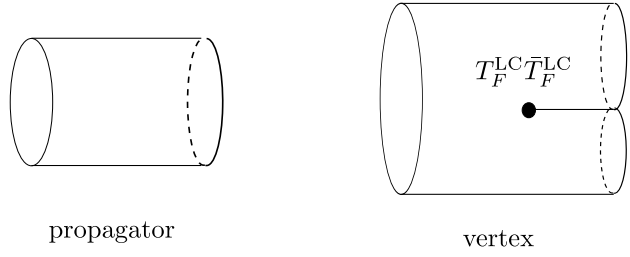} 
\par\end{centering}
\caption{The propagator and the vertex of the string field theory. \label{fig:The-propagator-and}}
\end{figure}

It is straightforward to calculate the amplitudes by the old-fashioned
perturbation theory starting from the action (\ref{eq:superHamiltonian})
and Wick rotate to Euclidean time. The propagator and the vertex are
given by the worldsheets depicted in figure \ref{fig:The-propagator-and},
where the left and right supercurrents $T_{\mathrm{F}}^{\mathrm{LC}}$,
$\bar{T}_{\mathrm{F}}^{\mathrm{LC}}$
of the transverse variables 
$X^{i}$, $\psi^{i}$, $\bar{\psi}^{i}$ $\left(i=1,\ldots,8\right)$
are inserted at the interaction points
of the three-string vertices. Each term in the expansion corresponds
to a light-cone gauge Feynman diagram for strings. 

\begin{figure}
\begin{centering}
\includegraphics[scale=1.4]{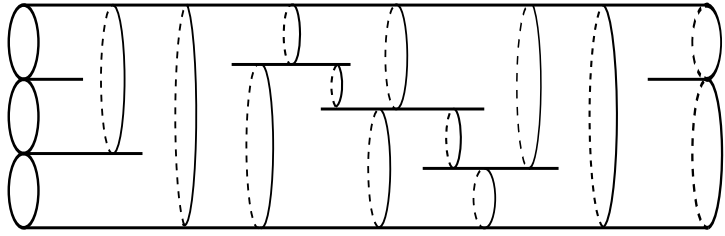} 
\par\end{centering}
\caption{A Feynman diagram of strings. \label{fig:A-string-diagram}}
\end{figure}

A typical light-cone gauge Feynman diagram for strings is depicted
in figure \ref{fig:A-string-diagram}. A Wick rotated $g$-loop $N$-string
diagram is conformally equivalent to an $N$ punctured genus $g$
Riemann surface $\Sigma$. A $g$-loop $N$-string amplitude is given
as an integral over the moduli space of $\Sigma$ as \cite{D'Hoker:1987pr,Aoki:1990yn}
\begin{equation}
\mathcal{A}_{N}^{(g)}=(ig_{s})^{2g-2+N}C\int[dT][\alpha d\theta][d\alpha]\,F_{N}^{(g)}~,\label{eq:ANg}
\end{equation}
where $\int[dT][\alpha d\theta][d\alpha]$ denotes the integration
over the moduli parameters and $C$ is the combinatorial factor. The
integrand $F_{N}^{(g)}$ is given as a path integral over the transverse
variables $X^{i}$, $\psi^{i}$, $\bar{\psi}^{i}$
on the light-cone diagram. A light-cone diagram consists of cylinders
which correspond to propagators of the closed string. On each cylinder
one can introduce a complex coordinate 
\begin{equation}
\rho=\tau+i\sigma\,,\label{eq:rho-coordinate}
\end{equation}
whose real part $\tau$ coincides with the Wick rotated light-cone
time $it$ and imaginary part $\sigma\sim\sigma+2\pi\alpha_{r}$ parametrizes
the closed string at each time. The $\rho$'s on the cylinders are
smoothly connected except at the interaction points and we get a complex
coordinate $\rho$ on $\Sigma$. The path integral on the light-cone
diagram is defined by using the metric
\begin{equation}
ds^{2}=d\rho d\bar{\rho}\,.\label{eq:canonical-metric}
\end{equation}
$\rho$ is not a good coordinate around the interaction points and
the punctures, and the metric (\ref{eq:canonical-metric}) is not
well-defined at these points. $F_{N}^{(g)}$ can be expressed in terms
of correlation functions defined with a metric $d\hat{s}^{2}=2\hat{g}_{z\bar{z}}dzd\bar{z}$
which is regular everywhere on the worldsheet, as 
\begin{eqnarray}
F_{N}^{(g)} & = & \left(2\pi\right)^{2}\delta\left(\sum_{r=1}^{N}p_{r}^{+}\right)\delta\left(\sum_{r=1}^{N}p_{r}^{-}\right)e^{-\frac{1}{2}\Gamma\left[\sigma;\hat{g}_{z\bar{z}}\right]}\nonumber \\
 &  & \times\sum_{\mathrm{spin}\ \mathrm{structure}}\int\left[dX^{i}d\psi^{i}d\bar{\psi}^{i}\right]_{\hat{g}_{z\bar{z}}}e^{-S^{\mathrm{LC}}\left[X^{i},\psi^{i},\bar{\psi}^{i}\right]}\nonumber \\
 &  & \hphantom{\sum_{\mathrm{spin}\ \mathrm{structure}}\int}
\times\prod_{I=1}^{2g-2+N}\left(\left|\partial^{2}\rho\left(z_{I}\right)\right|^{-\frac{3}{2}}T_{F}^{\mathrm{LC}}\left(z_{I}\right)\bar{T}_{F}^{\mathrm{LC}}\left(\bar{z}_{I}\right)\right)\prod_{r=1}^{N}V_{r}^{\mathrm{LC}}\left(Z_{r},\bar{Z}_{r}\right)\,.~~~\label{eq:superFN}
\end{eqnarray}
Here  $z$ is a complex coordinate of the Riemann surface
and the coordinate $\rho$ becomes a function $\rho (z)$ of $z$
(see e.g. \cite{Giddings:1986rf,D'Hoker:1988ta,Ishibashi:2013nma}).
$S^{\mathrm{LC}}\left[X^{i},\psi^{i},\bar{\psi}^{i}\right]$
denotes the worldsheet action of the transverse variables and the
path integral measure $\left[dX^{i}d\psi^{i}d\bar{\psi}^{i}\right]_{\hat{g}_{z\bar{z}}}$
is defined with the metric $d\hat{s}^{2}=2\hat{g}_{z\bar{z}}dzd\bar{z}$.
Since the integrand was defined by using the metric (\ref{eq:canonical-metric}),
we need the anomaly factor $e^{-\frac{1}{2}\Gamma\left[\sigma;\hat{g}_{z\bar{z}}\right]}$,
where 
\begin{eqnarray}
 &  & \sigma=\ln\partial\rho\bar{\partial}\bar{\rho}-\ln\hat{g}_{z\bar{z}}\,,\nonumber \\
 &  & \Gamma\left[\sigma;\hat{g}_{z\bar{z}}\right]=-\frac{1}{4\pi}\int dz\wedge d\bar{z}\sqrt{\hat{g}}\left(\hat{g}^{ab}\partial_{a}\sigma\partial_{b}\sigma+2\hat{R}\sigma\right)\,.\label{eq:Gammasigma}
\end{eqnarray}
$z_{I}\,\left(I=1,\cdots,2g-2+N\right)$ denote the $z$-coordinates
of the interaction points of the light-cone gauge Feynman diagram.
$V_{r}^{\mathrm{LC}}$ denotes the vertex operator for the $r$-th
external line inserted at $z=Z_{r}$ $(r=1,\ldots,N)$. The right
hand side of (\ref{eq:superFN}) does not depend on the choice of
$\hat{g}_{z\bar{z}}$. 

As was demonstrated in \cite{Ishibashi2017b}, if all the external
lines are in the (NS,NS) sector and the spin structure for the left
and right fermions are both even, the term in the sum in (\ref{eq:superFN})
can be recast into a conformal gauge expression:
\begin{eqnarray}
 &  & \int\left[dX^{\mu}d\psi^{\mu}d\bar{\psi}^{\mu}dbd\bar{b}dcd\bar{c}d\beta d\bar{\beta}d\gamma d\bar{\gamma}\right]_{\hat{g}_{z\bar{z}}}e^{-S^{\mathrm{tot}}}\nonumber \\
 &  & \qquad\times\prod_{K=1}^{6g-6+2N}\left[\oint_{C_{K}}\frac{dz}{\partial\rho}b_{zz}+\varepsilon_{K}\oint_{\bar{C}_{K}}\frac{d\bar{z}}{\bar{\partial}\bar{\rho}}b_{\bar{z}\bar{z}}\right]\prod_{I=1}^{2g-2+N}\left[X\left(z_{I}\right)\bar{X}\left(\bar{z}_{I}\right)\right]\nonumber \\
 &  & \qquad\times\prod_{r=1}^{N}V_{r}^{\left(-1,-1\right)}(Z_{r},\bar{Z}_{r})\,.\label{eq:conformal}
\end{eqnarray}
Here
$S^{\mathrm{tot}}$ denotes the worldsheet action for the variables
$X^{\mu},\psi^{\mu},\bar{\psi}^{\mu}$ ($\mu=+,-,1,\ldots,8$), ghosts
and superghosts, 
\begin{equation}
X\left(z\right)=\left[c\partial\xi-e^{\phi}T_{F}+\frac{1}{4}\partial b\eta e^{2\phi}+\frac{1}{4}b\left(2\partial\eta e^{2\phi}+\eta\partial e^{2\phi}\right)\right](z)\label{eq:PCO}
\end{equation}
is the picture changing operator (PCO), $\bar{X}\left(\bar{z}\right)$
is its antiholomorphic counterpart and $T_{F}$ denotes the supercurrent
for $\partial X^{\mu},\psi^{\mu}$. The contours $C_{K}$ and $\varepsilon_{K}=\pm1$
are chosen so that the antighost insertions correspond to the moduli
parameters for the light-cone amplitudes. 
The vertex operator $V_{r}^{\left(-1,-1\right)}(Z_{r},\bar{Z}_{r})$
is defined as
\begin{equation}
V_{r}^{\left(-1,-1\right)}(Z_{r},\bar{Z}_{r})\equiv c\bar{c}e^{-\phi-\bar{\phi}}V_{r}^{\mathrm{DDF}}(Z_{r},\bar{Z}_{r})\,.\label{eq:-1-1}
\end{equation}
$V_{r}^{\mathrm{DDF}}(Z_{r},\bar{Z}_{r})$ is the supersymmetric DDF
vertex operator given by 
\begin{equation}
V_{r}^{\mathrm{DDF}}(Z_{r},\bar{Z}_{r})=A_{-n_{1}}^{i_{1}(r)}\cdots\bar{A}_{-\bar{n}_{1}}^{\bar{i}_{1}(r)}\cdots B_{-s_{1}}^{j_{1}(r)}\cdots\bar{B}_{-\bar{s}_{1}}^{\bar{j}_{1}(r)}\cdots e^{-ip_{r}^{+}X^{-}-i\left(p_{r}^{-}-\frac{\mathcal{N}_{r}}{p_{r}^{+}}\right)X^{+}+ip_{r}^{i}X^{i}}(Z_{r},\bar{Z}_{r})~,\label{eq:superVDDF}
\end{equation}
with the DDF operators $A_{-n}^{i(r)},B_{-s}^{j(r)}$ for the $r$-th
string defined as
\begin{eqnarray}
A_{-n}^{i(r)} & = & \oint_{Z_{r}}\frac{d\mathbf{z}}{2\pi i}iD\mathcal{X}^{i}e^{-i\frac{n}{p_{r}^{+}}\mathcal{X}_{L}^{+}}(\mathbf{z})~,
\nonumber \\
B_{-s}^{i(r)} & = & \oint_{Z_{r}}\frac{d\mathbf{z}}{2\pi i}\frac{D\mathcal{X}^{+}}{\left(ip_{r}^{+}\partial\mathcal{X}^{+}\right)^{\frac{1}{2}}}D\mathcal{X}^{i}e^{-i\frac{s}{p_{r}^{+}}\mathcal{X}_{L}^{+}}(\mathbf{z})~,
\end{eqnarray}
and $\bar{A}_{-n}^{i(r)},\bar{B}_{-s}^{i(r)}$ are similarly given
for the antiholomorphic sector. Here we use the notation
\begin{eqnarray}
\mathcal{N}_{r} & \equiv & \sum_{k}n_{k}+\sum_{l}s_{l}=\sum_{\bar{k}}\bar{n}_{\bar{k}}+\sum_{\bar{l}}\bar{s}_{\bar{l}}\,,
\nonumber \\
\mathcal{X}^{\mu}\left(\mathbf{z},\bar{\mathbf{z}}\right) & = & X^{\mu}\left(z,\bar{z}\right)+i\theta\psi^{\mu}\left(z\right)+i\bar{\theta}\bar{\psi}^{\mu}\left(\bar{z}\right)+\theta\bar{\theta}F^{\mu}\,,
\nonumber \\
D & \equiv & \frac{\partial}{\partial\theta}+\theta\frac{\partial}{\partial z}
\,.
\end{eqnarray}
$\mathbf{z}=\left(z,\theta\right)$ denotes the superspace coordinate on the worldsheet
and $\mathcal{X}_{L}^{+}$ denotes the left-moving part of the superfield $\mathcal{X}^{+}$.\footnote{Although $\mathcal{X}_{L}^{+}$ is not a well-defined quantity, it
is used as a short-hand notation to express the vertex operator (\ref{eq:superVDDF}),
which is well-defined. } We take the vertex operators to satisfy the on-shell condition
\begin{equation}
\frac{1}{2}\left(-2p_{r}^{+}p_{r}^{-}+p_{r}^{i}p_{r}^{i}\right)+\mathcal{N}_{r}=\frac{1}{2}~.
\end{equation}
$V_{r}^{\mathrm{DDF}}(Z_{r},\bar{Z}_{r})$ turns out to be a weight
$\left(\frac{1}{2},\frac{1}{2}\right)$ primary field made from $X^{\mu},\psi^{\mu},\bar{\psi}^{\mu}$.
Therefore $V_{r}^{\left(-1,-1\right)}(Z_{r},\bar{Z}_{r})$ is an on-shell
vertex operator in $\left(-1,-1\right)$ picture. It is easy to see
that the expression of the amplitude (\ref{eq:ANg}) given as an integral
of (\ref{eq:conformal}) is BRST invariant. 

One way to derive the expression (\ref{eq:conformal}) is as follows
\cite{Ishibashi2017b}. Using a nilpotent fermionic charge, it is
possible to show that the right hand side of (\ref{eq:conformal})
is equal to 
\begin{eqnarray}
 &  & \int\left[dX^{\mu}d\psi^{\mu}d\bar{\psi}^{\mu}dbd\bar{b}dcd\bar{c}d\beta d\bar{\beta}d\gamma d\bar{\gamma}\right]_{\hat{g}_{z\bar{z}}}e^{-S^{\mathrm{tot}}}\nonumber \\
 &  & \qquad\times\prod_{K=1}^{6g-6+2N}\left[\oint_{C_{K}}\frac{dz}{\partial\rho}b_{zz}+\varepsilon_{K}\oint_{\bar{C}_{K}}\frac{d\bar{z}}{\bar{\partial}\bar{\rho}}b_{\bar{z}\bar{z}}\right]\prod_{I=1}^{2g-2+N}\left[e^{\phi}T_{\mathrm{F}}^{\mathrm{LC}}\left(z_{I}\right)e^{\bar{\phi}}\bar{T}_{\mathrm{F}}^{\mathrm{LC}}\left(\bar{z}_{I}\right)\right]\nonumber \\
 &  & \qquad\times\prod_{r=1}^{N}\left[c\bar{c}e^{-\phi-\bar{\phi}}V_{r}^{\mathrm{DDF}}(Z_{r},\bar{Z}_{r})\right]\,.\label{eq:conformal2}
\end{eqnarray}
In this form, the path integral factorizes into the contributions
from $X^{\mu},\psi^{\mu},\bar{\psi}^{\mu}$, ghosts and superghosts.
Each of these contributions is calculated by taking $\hat{g}_{z\bar{z}}$
to be the Arakelov metric $g_{z\bar{z}}^{\mathrm{A}}$ \cite{arakelov}.
In the matter sector, integration over the longitudinal variables
yields
\begin{eqnarray}
\lefteqn{
 \int\left[dX^{\pm}d\psi^{\pm}d\bar{\psi}^{\pm}
     \right]_{g^{\mathrm{A}}_{z\bar{z}}}
   e^{-S^{\pm}}
   \prod_{r=1}^{N}V_{r}^{\mathrm{DDF}}(Z_{r},\bar{Z}_{r})
} \nonumber \\
 & =& 
(2\pi)^2 \delta \left( \sum_{r=1}^{N}p^{-}_{r} \right)
\delta \left( \sum_{r=1}^{N} p^{+}_{r} \right)
    Z_{X^{\pm}} \left[ g^{\mathrm{A}}_{z\bar{z}} \right]
    Z_{\psi^{\pm}} \left[ g^{\mathrm{A}}_{z\bar{z}} \right]
 \prod_{r=1}^{N} \left[
       \frac{1}{\alpha_{r}}
       e^{-\mathop{\mathrm{Re}}\bar{N}_{00}^{rr}}
       V_{r}^{\mathrm{LC}}(Z_{r},\bar{Z}_{r})
    \right]\,,
\nonumber \\
\label{eq:Xpm}
\end{eqnarray}
where 
\begin{eqnarray}
Z_{X^{\pm}} \left[ g_{z\bar{z}} \right]
& = & \left( \frac{\det^{\prime}
                    \left( -g^{z\bar{z}}\partial_{z}\partial_{\bar{z}}
                    \right)}
                   {\int d^{2}z\sqrt{g}}
       \right)^{-1}\,,\nonumber \\
Z_{\psi^{\pm}} \left[ g_{z\bar{z}} \right] 
& = & \left(\frac{\det^{\prime}
                   \left(-g^{z\bar{z}}\partial_{z}\partial_{\bar{z}}
                   \right)}
                 {\det \mathop{\mathrm{Im}}\Omega \int d^{2}z\sqrt{g}}
      \right)^{-\frac{1}{2}}
\vartheta[\alpha_{\mathrm{L}}]\left(0\right)\vartheta[\alpha_{\mathrm{R}}]\left(0\right)^{\ast}\,,\label{eq:psipm}\\
\bar{N}_{00}^{rr} & = & \lim_{z\to Z_{r}}\left[\frac{\rho(z_{I^{(r)}})-\rho(z)}{\alpha_{r}}+\ln(z-Z_{r})\right]\,.\nonumber 
\end{eqnarray}
Here $\vartheta[\alpha]$ denotes the theta function with characteristic
$\alpha$ and  $\Omega$ is the period matrix. $\alpha_{\mathrm{L}}$ and $\alpha_{\mathrm{R}}$ denote
the characteristics corresponding to the spin structures of the left-
and right-moving fermions respectively. 
$Z_{X^{\pm}}\left[ g_{z\bar{z}} \right]$ and 
$Z_{\psi^{\pm}}\left[ g_{z\bar{z}} \right]$
are respectively the partition functions of the free variables $X^{\pm}$ 
and $\psi^{\pm},\bar{\psi}^{\pm}$ on the worldsheet endowed with
the metric $ds^2 = 2g_{z\bar{z}} dz d\bar{z}$. 
$z_{I^{\left(r\right)}}$ denotes the interaction point
at which the $r$-th string interacts. The contributions from the
ghosts and superghosts are given as 
\begin{eqnarray}
 &  & \int\left[dbd\bar{b}dcd\bar{c}\right]_{g_{z\bar{z}}^{\mathrm{A}}}e^{-S^{bc}}\prod_{r=1}^{N}c\bar{c}(Z_{r},\bar{Z}_{r})\prod_{K=1}^{6g-6+2N}\left[\oint_{C_{K}}\frac{dz}{\partial\rho}b_{zz}+\varepsilon_{K}\oint_{\bar{C}_{K}}\frac{d\bar{z}}{\bar{\partial}\bar{\rho}}b_{\bar{z}\bar{z}}\right]\nonumber \\
 &  & \qquad
=\left(Z_{X^{\pm}} \left[g^{\mathrm{A}}_{z\bar{z}} \right]
 \right)^{-1}
 e^{-\Gamma\left[\sigma; g_{z\bar{z}}^{\mathrm{A}}\right]}
 \prod_{r=1}^{N}\left(\alpha_{r}e^{2\mathop{\mathrm{Re}}\bar{N}_{00}^{rr}}
                \right)\,,\label{eq:bc}\\
 &  & \int\left[d\beta d\gamma d\bar{\beta}d\bar{\gamma}
          \right]_{g^{\mathrm{A}}_{z\bar{z}}}
     e^{-S_{\beta\gamma}}
     \prod_{I=1}^{2g-2+N}
         \left[ e^{\phi}\left(z_{I}\right)
                e^{\bar{\phi}}\left(\bar{z}_{I}\right)
         \right]
    \prod_{r=1}^{N}
          \left[ e^{-\phi}\left(Z_{r}\right)
                 e^{-\bar{\phi}}\left(\bar{Z}_{r}\right)
          \right]
\nonumber \\
 &  & \qquad
   =\left( Z_{\psi^{\pm}} \left[ g^{\mathrm{A}}_{z\bar{z}} \right]
    \right)^{-1}
    e^{\frac{1}{2}\Gamma\left[\sigma; g_{z\bar{z}}^{\mathrm{A}}\right]}
    \prod_{r=1}^{N}e^{-\mathop{\mathrm{Re}}\bar{N}_{00}^{rr}}
    \prod_{I=1}^{2g-2+N}\left|\partial^{2}\rho\left(z_{I}\right)
                        \right|^{-\frac{3}{2}}\,.
\label{eq:betagamma}
\end{eqnarray}
Substituting eqs.(\ref{eq:Xpm}), (\ref{eq:bc}), (\ref{eq:betagamma})
into (\ref{eq:conformal2}), we can easily see that (\ref{eq:conformal})
is equal to (\ref{eq:superFN}). 

\subsection{Dimensional regularization\label{subsec:Dimensional-regularization}}

The amplitudes of superstring theory were calculated using the first-quantized
formalism in \cite{Verlinde1987b} in which an expression using the
PCO's was given. The expression (\ref{eq:conformal}) is a special
case of the one in \cite{Verlinde1987b}, where the PCO's are placed
at the interaction points of the light-cone Feynman diagram.\footnote{Notice that in the light-cone setup, the positions of the PCO's have
the fixed coordinate in the coordinate patch on the surface and we
do not need $\partial\xi$ terms.} 

Unfortunately, the amplitude (\ref{eq:ANg}) given as an integral
of (\ref{eq:conformal}) or (\ref{eq:superFN}) is not well-defined.
(\ref{eq:superFN}) diverges when some of the interaction points collide,
because $T_{\mathrm{F}}^{\mathrm{LC}}(z)$ has the OPE
\begin{equation}
T_{\mathrm{F}}^{\mathrm{LC}}(z_{I})T_{\mathrm{F}}^{\mathrm{LC}}(z_{J})\sim\frac{2}{\left(z_{I}-z_{J}\right)^{3}}+\cdots\,,
\end{equation}
which makes the integral (\ref{eq:ANg}) ill-defined. This kind of
divergence is called the contact term divergence. 

Accordingly, the conformal gauge expression (\ref{eq:conformal})
suffers from the so-called spurious singularity. The holomorphic part
of the correlation function of the superghost system has the form
\begin{eqnarray}
 &  & \left\langle \prod_{I=1}^{2g-2+N}e^{\phi}\left(z_{I}\right)\prod_{r=1}^{N}e^{-\phi}\left(Z_{r}\right)\right\rangle 
\nonumber \\
 &  & \quad\sim\left[\vartheta[\alpha_{\mathrm{L}}]\left(-\sum_{r}\int_{P_{0}}^{Z_{r}}\omega+\sum_{I}\int_{P_{0}}^{z_{I}}\omega-2\int_{P_{0}}^{\bigtriangleup}\omega\right)\right]^{-1} 
\nonumber \\
 &  & \hphantom{\quad\sim}
\quad \times\frac{\prod_{I,r}E\left(z_{I},Z_{r}\right)}{\prod_{I<J}E\left(z_{I},z_{J}\right)\prod_{r<s}E\left(Z_{r},Z_{s}\right)}\frac{\prod_{r}\sigma^{2}\left(Z_{r}\right)}{\prod_{I}\sigma^{2}\left(z_{I}\right)}\,.
\end{eqnarray}
Here $\omega$ is the canonical basis of the holomorphic 1-forms,
$\bigtriangleup$ is the Riemann class, $E(z,w)$ is the prime form
of the surface and $\sigma\left(z\right)$ is a holomorphic $\frac{g}{2}$
form with no zeros or poles. The base point $P_{0}$ is an arbitrary
point on the surface.\footnote{For the mathematical background relevant for string perturbation theory,
we refer the reader to \cite{D'Hoker:1988ta}.} This correlation function diverges when
\begin{enumerate}
\item Some of $z_{I}$ collide. 
\item $\vartheta[\alpha_{\mathrm{L}}]\left(-\sum_{r}\int_{P_{0}}^{Z_{r}}\omega+\sum_{I}\int_{P_{0}}^{z_{I}}\omega-2\int_{P_{0}}^{\bigtriangleup}\omega\right)=0$.
\end{enumerate}
It also diverges when some of $Z_{r}$ collide, but such singularities
are at the boundary of moduli space of the punctured Riemann surface.
The singularities given above are called the spurious singularities.
The first type of singularity corresponds to the contact term divergence
mentioned above. The second type of singularity is due to existence
of zero modes of $\gamma$. Singularities of this kind do not arise
in our case. Since $Z_{r}\,\left(r=1,\ldots N\right)$ and $z_{I}\,\left(I=1,\ldots,2g-2+N\right)$
are the poles and the zeros of the meromorphic one-form $\partial\rho\left(z\right)dz$
respectively, $\sum_{I}z_{I}-\sum_{r}Z_{r}$ is a canonical divisor
on the surface. Therefore we obtain
\begin{equation}
-\sum_{r=1}^{N}\int_{P_{0}}^{Z_{r}}\omega+\sum_{I=1}^{2g-2+N}\int_{P_{0}}^{z_{I}}\omega\equiv2\int_{P_{0}}^{\bigtriangleup}\omega\qquad\pmod{\mathbb{Z}^{g}+\mathbb{Z}^{g}\Omega}\,,\label{eq:canonical-divisor}
\end{equation}
where $\Omega$ is the period matrix. This yields 
\begin{equation}
\left[\vartheta[\alpha_{\mathrm{L}}]\left(-\sum_{r}\int_{P_{0}}^{Z_{r}}\omega+\sum_{I}\int_{P_{0}}^{z_{I}}\omega-2\int_{P_{0}}^{\bigtriangleup}\omega\right)\right]^{-1}=\left[\vartheta[\alpha_{\mathrm{L}}]\left(0\right)\right]^{-1}\,,
\end{equation}
which is included in the factor $\left(Z_{\psi^{\pm}}\right)^{-1}$
in (\ref{eq:betagamma}). $\left[\vartheta[\alpha_{\mathrm{L}}]\left(0\right)\right]^{-1}$
may become singular at some points in the moduli space, but the $\left[\vartheta[\alpha_{\mathrm{L}}]\left(0\right)\right]^{-1}$
cancels the factor $\vartheta[\alpha_{\mathrm{L}}]\left(0\right)$
from the partition function $Z_{\psi^{\pm}}$ of $\psi^{\pm}$ and
the whole amplitude is free from this type of singularity. 

Therefore, in order to make the amplitudes given in the previous subsection
well-defined, we should deal with the contact term divergences. In
our previous works, we employ the dimensional regularization to do
so. Let us summarize the results:
\begin{itemize}
\item One can formulate the light-cone gauge superstring field theory in
$d\ne10$ dimensional space time. The amplitudes are given in the
form (\ref{eq:ANg}) with 
\begin{eqnarray}
F_{N}^{(g)} & = & \left(2\pi\right)^{2}\delta\left(\sum_{r=1}^{N}p_{r}^{+}\right)\delta\left(\sum_{r=1}^{N}p_{r}^{-}\right)e^{-\frac{d-2}{16}\Gamma\left[\sigma;\hat{g}_{z\bar{z}}\right]}\nonumber \\
 &  & \times\sum_{\mathrm{spin}\ \mathrm{structure}}\int\left[dX^{i}d\psi^{i}d\bar{\psi}^{i}\right]_{\hat{g}_{z\bar{z}}}e^{-S^{\mathrm{LC}}\left[X^{i},\psi^{i},\bar{\psi}^{i}\right]}\nonumber \\
 &  & \hphantom{\sum_{\mathrm{spin}\ \mathrm{structure}}\int\left[\right]}\times\prod_{I=1}^{2g-2+N}\left(\left|\partial^{2}\rho\left(z_{I}\right)\right|^{-\frac{3}{2}}T_{F}^{\mathrm{LC}}\left(z_{I}\right)\bar{T}_{F}^{\mathrm{LC}}\left(\bar{z}_{I}\right)\right)\prod_{r=1}^{N}V_{r}^{\mathrm{LC}}\,.~~~
\label{eq:DRamp}
\end{eqnarray}
Taking $d$ to be large and negative, the factor $e^{-\frac{d-2}{16}\Gamma\left[\sigma;\hat{g}_{z\bar{z}}\right]}$
tame the contact term divergences. 
\item More generally we can regularize the divergences by taking the worldsheet
superconformal field theory to be the one with central charge $c\ne12$.
One convenient choice of the worldsheet theory is the one in a linear
dilaton background $\Phi=-iQX^{1}$, with a real constant $Q$. The
worldsheet action of $X^{1}$ and its fermionic partners $\psi^{1},\bar{\psi}^{1}$
on a worldsheet with metric $d\hat{s}^{2}=2\hat{g}_{z\bar{z}}dzd\bar{z}$
becomes 
\begin{eqnarray}
S\left[X^{1},\psi^{1},\bar{\psi}^{1};\hat{g}_{z\bar{z}}\right] & = & \frac{1}{8\pi}\int dz\wedge d\bar{z}\sqrt{\hat{g}}\left(\hat{g}^{ab}\partial_{a}X^{1}\partial_{b}X^{1}-2iQ\hat{R}X^{1}\right)\nonumber \\
 &  & \quad {}+\frac{1}{4\pi}\int dz\wedge d\bar{z}i\left(\psi^{1}\bar{\partial}\psi^{1}+\bar{\psi}^{1}\partial\bar{\psi}^{1}\right)\,,\label{eq:linaction}
\end{eqnarray}
The amplitude is expressed in the form (\ref{eq:DRamp}) with 
\begin{equation}
d=10-8Q^{2}\,.
\end{equation}
It was shown in \cite{Ishibashi2017d} that (with the Feynman $i\varepsilon$)
by taking $Q^{2}>10$, the amplitudes become finite. 
\item We can define the amplitudes as analytic functions of $Q^{2}$ and
take the limit $Q\to0$ to obtain those in $d=10$. In order to study
the limit, it is useful to recast the expression (\ref{eq:DRamp})
into the conformal gauge one \cite{Ishibashi2017b}
\begin{eqnarray}
 &  & \int\left[dX^{\mu}d\psi^{\mu}d\bar{\psi}^{\mu}
               dbd\bar{b}dcd\bar{c}d\beta d\bar{\beta}
               d\gamma d\bar{\gamma}\right]_{g_{z\bar{z}}^{\mathrm{A}}}
       e^{-S^{\mathrm{tot}}}
\nonumber \\
 &  & \qquad\times
\prod_{K=1}^{6g-6+2N} \left[
     \oint_{C_{K}}\frac{dz}{\partial\rho}b_{zz}
     +\varepsilon_{K}\oint_{\bar{C}_{K}}
        \frac{d\bar{z}}{\bar{\partial}\bar{\rho}}b_{\bar{z}\bar{z}}
     \right]   
\prod_{I=1}^{2g-2+N}\left[X\left(z_{I}\right)\bar{X}\left(\bar{z}_{I}\right)
                    \right]
\nonumber \\
& & \qquad \times
\prod_{r}
   e^{-\frac{iQ^{2}}{\alpha_{r}}\mathcal{X}^{+}}
   \left(\hat{\tilde{\mathbf{z}}}_{I^{\left(r\right)}},
         \hat{\tilde{\bar{\mathbf{z}}}}_{I^{\left(r\right)}}
   \right)
\prod_{r=1}^{N}\left[V_{r}^{\left(-1,-1\right)}(Z_{r},\bar{Z}_{r})
\right],~~\label{eq:DRconformal}
\end{eqnarray}
which looks quite similar to 
the critical case (\ref{eq:conformal}).\footnote{Notice that 
the expression here is different from the one 
in \cite{Ishibashi2017b} where the operators
$$
\oint_{z_{I^{\left(r\right)}}}\frac{d\mathbf{z}}{2\pi i}\mathcal{S}
    \left(\mathbf{z},Z_{r}\right)
\oint_{\bar{z}_{I^{\left(r\right)}}}\frac{d\bar{\mathbf{z}}}{2\pi i}
    \bar{\mathcal{S}}\left(\bar{\mathbf{z}},\bar{Z}_{r}\right)
$$
are inserted in place of 
$e^{-\frac{iQ^{2}}{\alpha_{r}} \mathcal{X}^{+}}
  \left(\hat{\tilde{\mathbf{z}}}_{I^{\left(r\right)}},
        \hat{\tilde{\bar{\mathbf{z}}}}_{I^{\left(r\right)}}\right)$.
The properties of operators of this kind with operator valued arguments
$\hat{\tilde{\mathbf{z}}}_{I},\hat{\tilde{\bar{\mathbf{z}}}}_{I}$
are explained in appendix \ref{sec:Operator-valued-coordinate}. } The crucial difference is that the worldsheet theory for the longitudinal
variables $X^{\pm},\psi^{\pm},\bar{\psi}^{\pm}$ is a superconformal
field theory called the supersymmetric $X^{\pm}$ CFT, which has the
central charge
\begin{equation}
c=3+12Q^{2}\,.
\end{equation}
With this CFT, we can construct a nilpotent BRST charge. Using this
expression, one can show that the amplitudes in the limit $Q\to0$
coincide with those given by the Sen-Witten prescription \cite{Sen2015},
if the latter exists. 
\end{itemize}

\subsection{The problems with odd spin structure}

The light-cone gauge amplitudes can be defined and calculated for
odd spin structure, and we get the expression (\ref{eq:ANg}) with
the integrand given by (\ref{eq:superFN}). The correlation functions
of free fermions on higher genus Riemann surfaces are given in appendix
\ref{sec:Correlation-functions-of}. Only the amplitudes with enough
fermions from the vertex operators and $T_{\mathrm{F}}$ insertions
to soak up the zero modes are nonvanishing. However, we have a problem in rewriting
the light-cone gauge expression (\ref{eq:superFN}) into the BRST
invariant one (\ref{eq:conformal}), if we proceed as in the previous
section. If $\alpha$ corresponds to an odd spin structure,
\begin{equation}
\vartheta[\alpha]\left(0\right)=0\,.
\end{equation}
 The correlation function (\ref{eq:betagamma}) of the $\beta,\gamma$
system diverges because it involves factors 
\begin{equation}
\left(\vartheta[\alpha]\left(0\right)\right)^{-1}\,,
\end{equation}
coming from $\left(Z_{\psi^{\pm}}\right)^{-1}$ on the right hand side of   (\ref{eq:betagamma}).
On the other hand, the partition function (\ref{eq:psipm}) of the
$\psi^{\pm}$ variables involves factors 
\begin{equation}
\vartheta[\alpha]\left(0\right)\,,
\end{equation}
which cancel the divergent contribution from the $\beta,\gamma$ system.
Therefore we need to make sense out of the combination 
$$
0\times\infty
$$
to obtain the BRST invariant expression corresponding to the light-cone
gauge amplitudes. 

\section{Odd spin structure\label{sec:Odd-spin-structure}}

The problem mentioned at the end of the previous section can be avoided
by considering the amplitudes with insertions of $\psi^{+},\psi^{-}$
and $\delta(\beta),\delta(\gamma)$. We would like to show that such
insertions can be realized in a BRST invariant way, if we consider
the conformal gauge amplitudes taking some of the vertex operators
to have $0$ or $-2$ picture, when all the external lines are in
the (NS,NS) sector. 

\subsection{Multiloop amplitudes}

Let us consider the case where the spin structure $\alpha_{\mathrm{L}}$
for the left-moving fermions is odd and $\alpha_{\mathrm{R}}$ for
the right-moving fermions is even. The case where $\alpha_{\mathrm{L}}$
is even and $\alpha_{\mathrm{R}}$ is odd or both of $\alpha_{\mathrm{L}}$
and $\alpha_{\mathrm{R}}$ are odd can be dealt with in the same
way. We would like to show that the term 
\begin{eqnarray}
 &  & \int\left[dX^{i}d\psi^{i}d\bar{\psi}^{i}\right]_{\hat{g}_{z\bar{z}}}e^{-S^{\mathrm{LC}}\left[X^{i},\psi^{i},\bar{\psi}^{i}\right]}\nonumber \\
 &  & \hphantom{\times\sum_{\mathrm{spin}\ \mathrm{structure}}\int\quad}\times\prod_{I=1}^{2g-2+N}\left(\left|\partial^{2}\rho\left(z_{I}\right)\right|^{-\frac{3}{2}}T_{F}^{\mathrm{LC}}\left(z_{I}\right)\bar{T}_{F}^{\mathrm{LC}}\left(\bar{z}_{I}\right)\right)\prod_{r=1}^{N}V_{r}^{\mathrm{LC}}\left(Z_{r},\bar{Z}_{r}\right)\,,\label{eq:oddLC}
\end{eqnarray}
in the sum in (\ref{eq:superFN}) corresponding to such a spin structure
can be recast into a conformal gauge expression 
\begin{eqnarray}
 &  & \int\left[dX^{\mu}d\psi^{\mu}d\bar{\psi}^{\mu}dbd\bar{b}dcd\bar{c}d\beta d\bar{\beta}d\gamma d\bar{\gamma}\right]_{g_{z\bar{z}}^{\mathrm{A}}}e^{-S^{\mathrm{tot}}}\nonumber \\
 &  & \qquad\times\prod_{K=1}^{6g-6+2N}\left[\oint_{C_{K}}\frac{dz}{\partial\rho}b_{zz}+\varepsilon_{K}\oint_{\bar{C}_{K}}\frac{d\bar{z}}{\bar{\partial}\bar{\rho}}b_{\bar{z}\bar{z}}\right]\prod_{I}\left[X\left(z_{I}\right)\bar{X}\left(\bar{z}_{I}\right)\right]\nonumber \\
 &  & \qquad\times V_{1}^{\left(-2,-1\right)}\left(Z_{1},\bar{Z}_{1}\right)V_{2}^{\left(0,-1\right)}\left(Z_{2},\bar{Z}_{2}\right)\prod_{r=3}^{N}\left[V_{r}^{\left(-1,-1\right)}(Z_{r},\bar{Z}_{r})\right]\,,\label{eq:oddBRST}
\end{eqnarray}
up to a numerical factor. As we will see, the expression (\ref{eq:oddBRST})
is well-defined and free from the combination $0\times\infty$. Here
$V_{r}^{\left(-1,-1\right)}(Z_{r},\bar{Z}_{r})$ is the $\left(-1,-1\right)$
picture vertex operator defined in (\ref{eq:-1-1}) and $V_{1}^{\left(-2,-1\right)}\left(Z_{1},\bar{Z}_{1}\right),\,V_{2}^{\left(0,-1\right)}\left(Z_{2},\bar{Z}_{2}\right)$
are given by 
\begin{eqnarray}
V_{1}^{\left(-2,-1\right)}\left(Z_{1},\bar{Z}_{1}\right) & = & -\frac{2}{p_{1}^{+}}c\bar{c}e^{-2\phi}e^{-\bar{\phi}}\psi^{+}V_{1}^{\mathrm{DDF}}\left(Z_{1},\bar{Z}_{1}\right)\,,\label{eq:-2-1}\\
V_{2}^{\left(0,-1\right)}\left(Z_{2},\bar{Z}_{2}\right) & = & \left[-c\bar{c}e^{-\bar{\phi}}\oint_{Z_{2}}\frac{dz}{2\pi i}T_{\mathrm{F}}\left(z\right)+\frac{1}{4}\bar{c}\gamma e^{-\bar{\phi}}\right]V_{2}^{\mathrm{DDF}}\left(Z_{2},\bar{Z}_{2}\right),
\label{eq:0-1}
\end{eqnarray}
which satisfy
\begin{eqnarray}
XV_{1}^{\left(-2,-1\right)}\left(Z_{1},\bar{Z}_{1}\right) & = & V_{1}^{\left(-1,-1\right)}(Z_{1},\bar{Z}_{1})\,,
\nonumber \\
XV_{2}^{\left(-1,-1\right)}(Z_{2},\bar{Z}_{2}) & = & V_{2}^{\left(0,-1\right)}\left(Z_{2},\bar{Z}_{2}\right)\,,
\nonumber \\
Q_{\mathrm{B}}V_{1}^{\left(-2,-1\right)}\left(Z_{1},\bar{Z}_{1}\right) & = & 0\,,
\end{eqnarray}
where $X$ is the picture changing operator (\ref{eq:PCO}) and $Q_{\mathrm{B}}$
denotes the BRST charge (\ref{eq:BRSTcharge}). Since $V^{\mathrm{DDF}}\left(Z,\bar{Z}\right)$
is expressed as (\ref{eq:superVDDF}), $V_{2}^{\left(0,-1\right)}\left(Z_{2},\bar{Z}_{2}\right)$
can be rewritten as
\begin{eqnarray}
\lefteqn{
V_{2}^{\left(0,-1\right)}\left(Z_{2},\bar{Z}_{2}\right)
} \nonumber \\
 & & =  -c\bar{c}e^{-\bar{\phi}}A_{-n_{1}}^{i_{1}(2)}\cdots\bar{A}_{-\bar{n}_{1}}^{\bar{i}_{1}(2)}\cdots B_{-s_{1}}^{j_{1}(2)}\cdots\bar{B}_{-\bar{s}_{1}}^{\bar{j}_{1}(2)}\cdots
\nonumber \\
 &  & \hphantom{-c\bar{c}e^{-\bar{\phi}}}
\quad  \times\frac{1}{2}\left[p_{2}^{+}
 \psi^{-}+\left(p_{2}^{-}-\frac{\mathcal{N}_{2}}{p_{2}^{+}}\right)\psi^{+}-p_{2}^{i}\psi^{i}\right]e^{-ip_{2}^{+}X^{-}-i\left(p_{2}^{-}-\frac{\mathcal{N}_{2}}{p_{2}^{+}}\right)X^{+}+ip_{2}^{i}X^{i}}
\nonumber \\
 &  & \qquad{}+\frac{1}{4}\bar{c}\gamma e^{-\bar{\phi}}V_{2}^{\mathrm{DDF}}\left(Z_{2},\bar{Z}_{2}\right)
\nonumber \\
 & & =  -\frac{1}{2}c\bar{c}e^{-\bar{\phi}}p_{2}^{+}:\psi^{-}V_{2}^{\mathrm{DDF}}\left(Z_{2},\bar{Z}_{2}\right):+\cdots\,,
\end{eqnarray}
where the ellipses in the last line denote the terms which do not
involve $\psi^{-}$.

$V_{1}^{\left(-2,-1\right)}\left(Z_{1},\bar{Z}_{1}\right),\,V_{2}^{\left(0,-1\right)}\left(Z_{2},\bar{Z}_{2}\right)$
can be considered to be the BRST invariant vertex operators in $\left(-2,-1\right),\,\left(0,-1\right)$
pictures respectively. It is straightforward to define vertex operators
$V^{\left(-1,-2,\right)},V^{\left(-1,0\right)},$ or $V^{(-2,-2)},V^{(0,0)}$
which can be used to express the amplitudes for the cases of the other
spin structures mentioned above. 

It is possible to show that (\ref{eq:oddBRST}) is equal to 
\begin{eqnarray}
 &  & \int\left[dX^{\mu}d\psi^{\mu}d\bar{\psi}^{\mu}dbd\bar{b}dcd\bar{c}d\beta d\bar{\beta}d\gamma d\bar{\gamma}\right]_{g_{z\bar{z}}^{\mathrm{A}}}e^{-S^{\mathrm{tot}}}\nonumber \\
 &  & \qquad\times\prod_{K=1}^{6g-6+2N}\left[\oint_{C_{K}}\frac{dz}{\partial\rho}b_{zz}+\varepsilon_{K}\oint_{\bar{C}_{K}}\frac{d\bar{z}}{\bar{\partial}\bar{\rho}}b_{\bar{z}\bar{z}}\right]
\prod_{I}\left[e^{\phi}T_{F}^{\mathrm{LC}}\left(z_{I}\right)e^{\bar{\phi}}\bar{T}_{F}^{\mathrm{LC}}\left(\bar{z}_{I}\right)\right]\nonumber \\
 &  & \qquad\times V_{1}^{\left(-2,-1\right)}\left(Z_{1},\bar{Z}_{1}\right)V_{2}^{\left(0,-1\right)}\left(Z_{2},\bar{Z}_{2}\right)\prod_{r=3}^{N}\left[V_{r}^{\left(-1,-1\right)}(Z_{r},\bar{Z}_{r})\right]\,.\label{eq:oddBRST2}
\end{eqnarray}
A poof of this fact can be found in
appendix \ref{subsec:A-proof-of}. Therefore, in order to show that
(\ref{eq:oddLC}) is proportional to (\ref{eq:oddBRST}), we evaluate
(\ref{eq:oddBRST2}) and prove that it is proportional to (\ref{eq:oddLC}).
In (\ref{eq:oddBRST2}), we can replace $V_{2}^{\left(0,-1\right)}\left(Z_{2},\bar{Z}_{2}\right)$
by
\begin{equation}
-\frac{1}{2}c\bar{c}e^{-\bar{\phi}}p_{2}^{+}:\psi^{-}V_{2}^{\mathrm{DDF}}:\left(Z_{2},\bar{Z}_{2}\right)\,,\label{eq:V_20-1}
\end{equation}
because only this term can soak up the zero mode of $\psi^{-}$. After
such a replacement, (\ref{eq:oddBRST2}) factorizes into contributions
from the ghosts, superghosts, longitudinal modes and the transverse
modes. 
The parts of the transverse variables and ghosts are the same as
those in (\ref{eq:conformal2})
and we can use (\ref{eq:bc}) to evaluate the latter.
In this case,
the correlation function of the longitudinal variables is
modified from (\ref{eq:Xpm}) into
\begin{eqnarray}
\lefteqn{
 \int \left[dX^{\pm}d\psi^{\pm}d\bar{\psi}^{\pm}
      \right]_{g^{\mathrm{A}}_{z\bar{z}}}
   e^{-S^{\pm}}
    \, \psi^{+} \left(Z_{1} \right) \psi^{-} \left( Z_{2} \right)
   \prod_{r=1}^{N}V_{r}^{\mathrm{DDF}}(Z_{r},\bar{Z}_{r})
} \nonumber \\
 &  & =
(2\pi)^2 \delta \left( \sum_{r=1}^{N}p^{-}_{r} \right)
\delta \left( \sum_{r=1}^{N} p^{+}_{r} \right)
    Z_{X^{\pm}} \left[ g^{\mathrm{A}}_{z\bar{z}} \right]
   \left( \frac{\det'\left( -g^{\mathrm{A} z \bar{z}}
                            \partial_{z} \partial_{\bar{z}} \right)}
                {\det \mathop{\mathrm{Im}} \Omega
                 \int d^2z \sqrt{g^{\mathrm{A}}}}
   \right)^{-\frac{1}{2}}
   \vartheta \left[ \alpha_{\mathrm{R}} \right] (0)^{\ast}
\nonumber \\
& & \qquad
\times h_{\alpha_{\mathrm{L}}} \left( Z_{1} \right)
       h_{\alpha_{\mathrm{L}}} \left( Z_{2} \right)
 \prod_{r=1}^{N} \left[
       \frac{1}{\alpha_{r}}
       e^{-\mathop{\mathrm{Re}}\bar{N}_{00}^{rr}}
       V_{r}^{\mathrm{LC}}(Z_{r},\bar{Z}_{r})
    \right]\,,
\label{eq:Xpm-Lodd}
\end{eqnarray}
and that of the superghosts
is evaluated to be
\begin{eqnarray}
\lefteqn{
\int\left[d\beta d\gamma d\bar{\beta}d\bar{\gamma}
          \right]_{g^{\mathrm{A}}_{z\bar{z}}}
      e^{-S_{\beta\gamma}}
\, e^{-2\phi}\left(Z_{1}\right)
  \prod_{I} \left[e^{\phi}\left(z_{I}\right)
                  e^{\bar{\phi}}\left(\bar{z}_{I}\right)
            \right]
  \prod_{r=3}^{N}e^{-\phi}\left(Z_{r}\right)
  \prod_{r=1}^{N}e^{-\bar{\phi}}\left(\bar{Z}_{r}\right)
}  \nonumber \\
 & \propto &
 \left(\frac{\det^{\prime}\left(
             -g^{\mathrm{A} z\bar{z}}
             \partial_{z}\partial_{\bar{z}}\right)}
            {\det\mathop{\mathrm{Im}}\Omega
             \int d^{2}z\sqrt{g^{\mathrm{A}}}}
 \right)^{\frac{1}{2}}
\nonumber \\
 &  & 
\quad \times\left[\vartheta[\alpha_{L}]
    \left(-\sum_{r=1}^{N}\int_{P_{0}}^{Z_{r}}\omega
          +\sum_{I}\int_{P_{0}}^{z_{I}}\omega
          -2\int_{P_{0}}^{\triangle}+\int_{Z_{1}}^{Z_{2}}\omega
   \right)  \right]^{-1} 
\nonumber \\
& & 
\quad 
\times \left[ \vartheta[\alpha_{R}]
  \left(-\sum_{r=1}^{N}\int_{P_{0}}^{Z_{r}}\omega
        +\sum_{I}\int_{P_{0}}^{z_{I}}\omega
        -2\int_{P_{0}}^{\triangle}\omega\right)^{\ast}\right]^{-1}
\nonumber \\
 &  & 
\quad 
\times\left|
  \frac{\prod_{I,r}E\left(z_{I},Z_{r}\right)}
       {\prod_{I<J}E\left(z_{I},z_{J}\right)
         \prod_{r<s}E\left(Z_{r},Z_{s}\right)}
  \frac{\prod_{r}\sigma^{2}\left(Z_{r}\right)}
       {\prod_{I}\sigma^{2}\left(z_{I}\right)}
  \right|^{2}e^{-12S}\nonumber \\
 &  & 
\quad
\times\frac{\prod_{r=3}^{N}E(Z_{2},Z_{r})
             \prod_{I}E(z_{I},Z_{1})}
           {\prod_{I}E(z_{I},Z_{2})
             \prod_{r=3}^{N} E(Z_{1},Z_{r})}
  \cdot
     \frac{\sigma^{2}(Z_{1})}{\sigma^{2}(Z_{2})}
  \cdot 
     E\left(Z_{1},Z_{2}\right)
\nonumber \\
& \propto & 
  \frac{\alpha_{1}}{\alpha_{2}}
  \left( \frac{ \det^{\prime} \left(
                  -g^{\mathrm{A} z\bar{z}}\partial_{z}\partial_{\bar{z}}
                     \right)}
              {\det\mathop{\mathrm{Im}} \Omega
               \int d^{2}z \sqrt{g^{\mathrm{A}}}}
  \right)^{\frac{1}{2}}
  \left| 
    \frac{\prod_{I,r}E\left(z_{I},Z_{r}\right)}
         {\prod_{I<J}E\left(z_{I},z_{J}\right)
           \prod_{r<s}E\left(Z_{r},Z_{s}\right)}
    \frac{\prod_{r}\sigma^{2}\left(Z_{r}\right)}
         {\prod_{I}\sigma^{2}\left(z_{I}\right)}
  \right|^{2}e^{-12S}
\nonumber \\
& & 
\quad \times 
\frac{1}{h_{\alpha_{\mathrm{L}}}(Z_{1})h_{\alpha_{\mathrm{L}}}(Z_{2})
          \vartheta[\alpha_{\mathrm{R}}]\left(0\right)^{\ast}}
\nonumber \\
&=& \frac{\alpha_{1}}{\alpha_{2}}
  \left( \frac{ \det^{\prime} \left(
                  -g^{\mathrm{A} z\bar{z}}\partial_{z}\partial_{\bar{z}}
                     \right)}
              {\det\mathop{\mathrm{Im}} \Omega
               \int d^{2}z \sqrt{g^{\mathrm{A}}}}
  \right)^{\frac{1}{2}}
 e^{\frac{1}{2} \Gamma \left[ \sigma; g^{\mathrm{A}}_{z\bar{z}} \right]}
 \prod_{r} e^{-\mathop{\mathrm{Re}} \bar{N}^{rr}_{00}}
 \prod_{I} \left| \partial^2 \rho (z_{I}) \right|^{-\frac{3}{2}}
\nonumber \\
& & \quad \times 
\frac{1}{h_{\alpha_{\mathrm{L}}}(Z_{1})h_{\alpha_{\mathrm{L}}}(Z_{2})
          \vartheta[\alpha_{\mathrm{R}}]\left(0\right)^{\ast}}
\,.\label{eq:superghostodd}
\end{eqnarray}
Here $h_{\alpha_{\mathrm{L}}}(z)$ defined in (\ref{eq:halphaL})
is equal to the zero mode of spin $\frac{1}{2}$ left-moving fermion
with spin structure $\alpha_{\mathrm{L}}$. 
The explicit form of $S$ and its relation to $e^{-\Gamma\left[
\sigma; g^{\mathrm{A}}_{z\bar{z}} \right]}$
can be found in \cite{Ishibashi2017b}.
In the manipulations in
(\ref{eq:superghostodd}), we have used (\ref{eq:canonical-divisor})
and the following identities:
\begin{eqnarray}
E\left(Z_{2},Z_{1}\right) 
& = & \frac{\vartheta[\alpha_{\mathrm{L}}]
              \left(\int_{Z_{1}}^{Z_{2}}\omega\right)}
           {h_{\alpha_{\mathrm{L}}}(Z_{1})h_{\alpha_{\mathrm{L}}}(Z_{2})}
\,,\nonumber \\
\frac{\prod_{r=3}^{N}E(Z_{2},Z_{r})\prod_{I}E(z_{I},Z_{1})}
     {\prod_{I}E(z_{I},Z_{2})\prod_{r=3}^{N}E(Z_{1},Z_{r})}
 \cdot
\frac{\sigma^{2}(Z_{1})}{\sigma^{2}(Z_{2})}
 & = & -\frac{\alpha_{1}}{\alpha_{2}}\,.
\label{eq:alpha1alpha2}
\end{eqnarray}
(\ref{eq:alpha1alpha2}) is proved by observing 
\begin{equation}
\left|\partial\rho\left(z\right)\right|^{2}
=C \left|\sigma\left(z\right)\right|^{4}
   \left|\frac{\prod_{I}E\left(z,z_{I}\right)}
              {\prod_{r}E\left(z,Z_{r}\right)}
   \right|^{2}\,,
\end{equation}
where $C$ is a quantity independent of $z$. From this expression
we can derive 
\begin{eqnarray}
\frac{\alpha_{1}}{\alpha_{2}} 
& = & \lim_{z\to Z_{1},\,w\to Z_{2}}
       \frac{\left(z-Z_{1}\right)\partial\rho\left(z\right)}
            {\left(w-Z_{2}\right)\partial\rho\left(w\right)}
\nonumber \\
 & = & \lim_{z\to Z_{1},\,w\to Z_{2}}
       \frac{z-Z_{1}}{w-Z_{2}}
        \exp\left[\int_{w}^{z}du\,
                  \partial\ln\left|\partial\rho\left(u\right)
                              \right|^{2}
             \right]
\nonumber \\
 & = & -\frac{\prod_{r=3}^{N} E(Z_{2},Z_{r})
               \prod_{I}E(z_{I},Z_{1})}
              {\prod_{I}E(z_{I},Z_{2})\prod_{r=3}^{N} E(Z_{1},Z_{r})}
   \cdot  \frac{\sigma^{2}(Z_{1})}{\sigma^{2}(Z_{2})}
\,.
\end{eqnarray}
Combining eqs.(\ref{eq:bc}), (\ref{eq:Xpm-Lodd}) 
and (\ref{eq:superghostodd}),
it is straightforward to show that (\ref{eq:oddBRST2}) is proportional
to (\ref{eq:oddLC}). 

\subsection{Dimensional regularization}

The amplitudes given by the integral (\ref{eq:ANg}) with the integrand
of the form (\ref{eq:oddBRST}) is not well-defined, because of the
contact term divergences. In order to make them well-defined, we employ
the dimensional regularization illustrated 
in subsection \ref{subsec:Dimensional-regularization}.
The amplitudes are given in the form (\ref{eq:ANg}) with the integrand
(\ref{eq:DRamp}) with $d=10-8Q^2$. The light-cone gauge amplitudes are finite for
$Q^{2}>10$. As in the case of even spin structure, we can define
the amplitudes as analytic functions of $Q^{2}$ and take the limit
$Q\to0$ to obtain those in $d=10$. In order to study the limit,
we recast the light-cone gauge expression into a conformal gauge one.
The noncritical version of (\ref{eq:oddBRST}) is given as
\begin{eqnarray}
 &  & \int\left[dX^{\mu}d\psi^{\mu}d\bar{\psi}^{\mu}dbd\bar{b}dcd\bar{c}d\beta d\bar{\beta}d\gamma d\bar{\gamma}\right]_{g_{z\bar{z}}^{\mathrm{A}}}e^{-S^{\mathrm{tot}}}
\nonumber \\
 &  & \qquad\times
\prod_{K=1}^{6g-6+2N}\left[\oint_{C_{K}}\frac{dz}{\partial\rho}b_{zz}+\varepsilon_{K}\oint_{\bar{C}_{K}}\frac{d\bar{z}}{\bar{\partial}\bar{\rho}}b_{\bar{z}\bar{z}}\right]
\prod_{I}\left[X\left(z_{I}\right)\bar{X}\left(\bar{z}_{I}\right)\right]
\nonumber \\
& & \qquad \times
\prod_{r}e^{-\frac{iQ^{2}}{\alpha_{r}}\mathcal{X}^{+}}\left(\hat{\tilde{\mathbf{z}}}_{I^{\left(r\right)}},\hat{\tilde{\bar{\mathbf{z}}}}_{I^{\left(r\right)}}\right)
\nonumber \\
 &  & \qquad\times 
V_{1}^{\left(-2,-1\right)}\left(Z_{1},\bar{Z}_{1}\right)V_{2}^{\left(0,-1\right)}\left(Z_{2},\bar{Z}_{2}\right)\prod_{r=3}^{N}\left[V_{r}^{\left(-1,-1\right)}(Z_{r},\bar{Z}_{r})\right].\label{eq:ncBRSTodd}
\end{eqnarray}
Here the worldsheet theory of the longitudinal variables are taken
to be the supersymmetric $X^{\pm}$ CFT. We discuss the correlation
functions of the supersymmetric $X^{\pm}$ CFT for odd spin structures
in appendix \ref{subsec:Supersymmetric--CFT}. As is shown in appendix
\ref{subsec:A-proof-of}, this expression is equal to 
\begin{eqnarray}
 &  & \int\left[dX^{\mu}d\psi^{\mu}d\bar{\psi}^{\mu}dbd\bar{b}dcd\bar{c}d\beta d\bar{\beta}d\gamma d\bar{\gamma}\right]_{g_{z\bar{z}}^{\mathrm{A}}}e^{-S^{\mathrm{tot}}}\nonumber \\
 &  & \qquad\times\prod_{K=1}^{6g-6+2N}\left[\oint_{C_{K}}\frac{dz}{\partial\rho}b_{zz}+\varepsilon_{K}\oint_{\bar{C}_{K}}\frac{d\bar{z}}{\bar{\partial}\bar{\rho}}b_{\bar{z}\bar{z}}\right]
\prod_{I}\left[e^{\phi}T_{F}^{\mathrm{LC}}\left(z_{I}\right)e^{\bar{\phi}}\bar{T}_{F}^{\mathrm{LC}}\left(\bar{z}_{I}\right)\right]
\nonumber \\
& & \qquad \times 
\prod_{r}e^{-\frac{iQ^{2}}{\alpha_{r}}\mathcal{X}^{+}}\left(\hat{\tilde{\mathbf{z}}}_{I^{\left(r\right)}},\hat{\tilde{\bar{\mathbf{z}}}}_{I^{\left(r\right)}}\right)\nonumber \\
 &  & \qquad\times V_{1}^{\left(-2,-1\right)}\left(Z_{1},\bar{Z}_{1}\right)V_{2}^{\left(0,-1\right)}\left(Z_{2},\bar{Z}_{2}\right)\prod_{r=3}^{N}\left[V_{r}^{\left(-1,-1\right)}(Z_{r},\bar{Z}_{r})\right]\,.\label{eq:ncBRST2odd}
\end{eqnarray}
In (\ref{eq:ncBRST2odd}), we can replace $V_{2}^{\left(0,-1\right)}\left(Z_{2},\bar{Z}_{2}\right)$
by (\ref{eq:V_20-1}) for the same reason as that in the critical
case. With the replacement, the path integral (\ref{eq:ncBRST2odd})
can factorize into those of matter, ghosts and superghosts. For the
longitudinal variables, we get from (\ref{eq:superXpmCFTcorrodd})
\begin{eqnarray}
\lefteqn{
\int\left[d\mathcal{X}^{+}d\mathcal{X}^{-}
          \right]_{g^{\mathrm{A}}_{z\bar{z}}}
      e^{-S_{\mathrm{super}}^{\pm}\left[g^{\mathrm{A}}_{z\bar{z}}\right]}
\prod_{r=1}^{N}\left[V_{r}^{\mathrm{DDF}}(Z_{r},\bar{Z}_{r})e^{-\frac{iQ^{2}}{\alpha_{r}}X^{+}}\left(\hat{\tilde{\mathbf{z}}}_{I^{\left(r\right)}},\hat{\tilde{\bar{\mathbf{z}}}}_{I^{\left(r\right)}}\right)\right]\psi^{+}\left(Z_{1}\right)\psi^{-}\left(Z_{2}\right)
}\nonumber \\
 & = & (2\pi)^{2}\delta\left(\sum_{s}p_{s}^{-}\right)
                 \delta\left(\sum_{r}p_{r}^{+}\right)
Z_{X^{\pm}} \left[ g^{\mathrm{A}}_{z\bar{z}} \right]
\left(\frac{\det^{\prime}
              \left(-g^{\mathrm{A} z\bar{z}}\partial_{z}\partial_{\bar{z}}
              \right)}
            {\det \mathop{\mathrm{Im}} \Omega
             \int d^{2} z \sqrt{g^{\mathrm{A}}}}
\right)^{-\frac{1}{2}}
\vartheta[\alpha_{\mathrm{R}}]\left(0\right)^{\ast}
\nonumber \\
 &  & 
\qquad 
\times e^{\frac{Q^{2}}{2}\Gamma\left[\sigma;g^{\mathrm{A}}_{z\bar{z}}\right]}h_{\alpha_{\mathrm{L}}}\left(Z_{1}\right)h_{\alpha_{\mathrm{L}}}\left(Z_{2}\right)
\prod_{r}\left[\frac{1}{\alpha_{r}}e^{-\mathrm{Re}\bar{N}_{00}^{rr}}V_{r}^{\mathrm{LC}}\right]\,.\label{eq:longiodd}
\end{eqnarray}
Combining eqs.(\ref{eq:bc}), (\ref{eq:superghostodd}) and (\ref{eq:longiodd}),
it is straightforward to show that (\ref{eq:ncBRST2odd}) is proportional
to the light-cone gauge expression in (\ref{eq:DRamp})
with $d=10-8Q^2$. 

The expression (\ref{eq:ncBRSTodd}),
summed over spin structures and 
integrated over the moduli parameters, 
gives a BRST invariant expression of the amplitude. The operators
$X,\bar{X}$, 
$e^{-\frac{iQ^{2}}{\alpha_{r}}\mathcal{X}^{+}}
   \left(\hat{\tilde{\mathbf{z}}}_{I^{\left(r\right)}},
         \hat{\tilde{\bar{\mathbf{z}}}}_{I^{\left(r\right)}}\right)$, 
$V_{1}^{\left(-2,-1\right)}$, $V_{2}^{\left(0,-1\right)}$, 
$V_{r}^{\left(-1,-1\right)}$
are all BRST invariant and the BRST variations of the antighost insertions
yield total derivatives on moduli space. Since
(\ref{eq:ncBRSTodd}) coincides with (\ref{eq:DRamp}), 
the amplitude is finite for $Q^{2}>10$. 

We use the light-cone gauge expression of the amplitude for $Q^{2}>10$
to define it as an analytic function of $Q^{2}$, which is denoted
by $A^{\mathrm{LC}}\left(Q^{2}\right)$. We would like to see what
happens in the limit $Q\to0.$ 
The conformal gauge expression (\ref{eq:ncBRSTodd})
can be deformed to define the amplitudes following the Sen-Witten
prescription \cite{Sen2015a,Sen2015}. We can divide the moduli space
into patches and put the PCO's avoiding the spurious singularities
as was explained in \cite{Sen2015} and define the amplitude $A^{\mathrm{SW}}\left(Q^{2}\right)$.
Moving the locations of the PCO's, the amplitudes change by total
derivative terms in moduli space. Taking $Q^{2}$ big enough, these total derivative
terms do not contribute to the amplitudes, because the infrared divergences
are regularized. 
Therefore
$A^{\mathrm{SW}}\left(Q^{2}\right)$ coincides
with $A^{\mathrm{LC}}\left(Q^{2}\right)$ 
as an analytic function of $Q^{2}$. 
Since $A^{\mathrm{SW}}\left(Q^{2}\right)$
is free from the spurious singularities, it can be well-defined for
$Q^{2}<10$ and 
\begin{equation}
\lim_{Q\to0}A^{\mathrm{LC}}\left(Q^{2}\right)=A^{\mathrm{SW}}\left(0\right)\,,
\end{equation}
if the right hand side is well-defined. 

\section{Conclusions and discussions\label{sec:Conclusions-and-discussions}}

In this paper, we have shown that the Feynman amplitudes of the light-cone
gauge closed superstring field theory can be calculated using the
dimensional regularization technique, for higher genus Riemann surfaces
with odd spin structure, if the external lines are in the (NS,NS)
sector. In order to deal with the fermion zero modes peculiar to odd
spin structures, we need to change the pictures of the vertex operators
in the conformal gauge expression. We obtain the amplitudes in noncritical
dimensions which coincide with the ones defined by using the Sen-Witten
prescription. The amplitudes in the critical dimensions correspond
to the limit $d\to10$ or $Q\to0$, and the results coincide with
those given by the Sen-Witten prescription.

There are several things remain to be done. One is to check how the
amplitudes obtained by our procedure are related to the standard results in more detail.
In particular, we should study the conditionally convergent integrals
which appear in the Feynman amplitudes of superstrings. We expect
that our regularization makes the integrals well-defined but in a
way different from those in \cite{Witten2013a,Sen2015d}. Another
thing to be done is to generalize our results to the amplitudes with
external lines in the Ramond sector. With the correlation functions
involving spin fields given for example in \cite{Atick1987b}, it
will be straightforward to rewrite the light-cone gauge expression
into the conformal gauge one. These problems are left to future work. 

\section*{Acknowledgments}

N.I. would like to thank Ashoke Sen for useful comments. He would
also like to thank the organizers of ``Recent Developments on Light
Front'' at Arnold Sommerfeld Center for Theoretical Physics and ``SFT@HIT''
at Holon, especially Ivo Sachs, Ted Erler, Sebastian Konopka and Michael
Kroyter, for hospitality. This work was supported in part by Grant-in-Aid
for Scientific Research (C) (25400242) and (15K05063) from MEXT.

\appendix

\section{Operator valued coordinate\label{sec:Operator-valued-coordinate}}

It is convenient to introduce the operator valued coordinate $\hat{z}_{I}$
and its supersymmetric version $\hat{\tilde{\mathbf{z}}}_{I}$, in
order to express the conformal gauge form of the amplitudes for noncritical
dimensions. 

Let us first consider $\hat{z}_{I}$ which is defined in the bosonic
case. In the light-cone gauge setup, we consider the situation where
the variable $X^{+}\left(z,\bar{z}\right)$ possesses an expectation
value $-\frac{i}{2}\left(\rho\left(z\right)+\bar{\rho}\left(\bar{z}\right)\right)$.
Therefore we decompose it into the expectation value and the fluctuation
as 
\begin{equation}
X^{+}\left(z,\bar{z}\right)=-\frac{i}{2}\left(\rho\left(z\right)+\bar{\rho}\left(\bar{z}\right)\right)+\delta X^{+}\left(z,\bar{z}\right)\,.\label{eq:deltaX+}
\end{equation}
Roughly speaking, we define $\hat{z}_{I}$ to be an operator valued
coordinate which satisfies
\begin{equation}
\partial X^{+}\left(\hat{z}_{I}\right)=0\,.\label{eq:hatzI}
\end{equation}
Substituting (\ref{eq:deltaX+}) into (\ref{eq:hatzI}), we get 
\begin{equation}
-\frac{i}{2}\partial\rho\left(\hat{z}_{I}\right)+\partial\delta X^{+}\left(\hat{z}_{I}\right)=0\,.\label{eq:hatzI2}
\end{equation}
We take $\hat{z}_{I}$ so as to coincide with $z_{I}$ when $\delta X^{+}=0$.
Assuming that $\hat{z}_{I}$ is expanded in terms of the fluctuation
$\delta X^{+}$ as
\begin{equation}
\hat{z}_{I}=z_{I}+\sum_{n=1}^{\infty}\delta^{\left(n\right)}\hat{z}_{I}\,,\label{eq:hatzIexp}
\end{equation}
where $\delta^{\left(n\right)}z_{I}$ is at the $n$-th order in the
derivatives of $\delta X^{+}$, in principle we can obtain $\delta^{\left(n\right)}z_{I}$
if $\partial^{2}\rho\left(z_{I}\right)\ne0$. Lower order examples
are given by
\begin{eqnarray}
\delta^{\left(1\right)}\hat{z}_{I} & = & -\frac{2i}{\partial^{2}\rho}\partial\delta X^{+}\left(z_{I}\right)\,,
\nonumber \\
\delta^{\left(2\right)}\hat{z}_{I} & = & -\frac{4}{\left(\partial^{2}\rho\right)^{2}}\partial\delta X^{+}\partial^{2}\delta X^{+}\left(z_{I}\right)+\frac{2\partial^{3}\rho}{\left(\partial^{2}\rho\right)^{3}}\left(\partial\delta X^{+}\right)^{2}\left(z_{I}\right)\,.
\end{eqnarray}
In general $\delta^{\left(n\right)}z_{I}$ becomes a polynomial of
the derivatives of $\delta X^{+}$ at $z=z_{I}$. Quantities of order
$n$ with $n>N$ for some $N>0$ do not contribute to the correlation
functions we consider in this paper. $\hat{\bar{z}}_{I}$, which is
the antiholomorphic counterpart of $\hat{z}_{I}$, can be obtained
in the same way. 

The OPE of $\hat{z}_{I}$ with the energy-momentum tensor $T\left(z\right)$
comes from the contractions of $\partial X^{-}$ in $T\left(z\right)$
with $\partial^{k}\delta X^{+}\left(z_{I}\right)$ in $\hat{z}_{I}$.
Taking the OPE of (\ref{eq:hatzI}) with $T\left(z\right)$, we get
\begin{equation}
\begC1{T}\conC{\left(z\right)}\endC1{\hat{z}_{I}}\partial^{2}X^{+}\left(\hat{z}_{I}\right)+\frac{1}{\left(z-\hat{z}_{I}\right)^{2}}\partial X^{+}\left(\hat{z}_{I}\right)+\frac{1}{z-\hat{z}_{I}}\partial^{2}X^{+}\left(\hat{z}_{I}\right)=0\,.
\end{equation}
Using (\ref{eq:hatzI}) and the fact that $\partial^{2}X^{+}\left(\hat{z}_{I}\right)=-\frac{i}{2}\partial^{2}\rho\left(z_{I}\right)+\cdots$
is invertible perturbatively, we obtain
\begin{equation}
\begC1{T}\conC{\left(z\right)}\endC1{\hat{z}_{I}}\sim-\frac{1}{z-\hat{z}_{I}}\,.\label{eq:OPETzIhat}
\end{equation}
Expanding $\hat{z}_{I}$ as in (\ref{eq:hatzIexp}), we can see that
the right hand side of (\ref{eq:OPETzIhat}) involves poles of arbitrarily
high order at $z=z_{I}$. Only finite number of them are relevant
in the correlation functions (\ref{eq:DRconformal}), (\ref{eq:ncBRSTodd})
and (\ref{eq:ncBRST2odd}). 

With the OPE (\ref{eq:OPETzIhat}) and its antiholomorphic version,
we can show the OPE's
\begin{eqnarray}
T\left(z\right)e^{i\alpha X^{+}}\left(\hat{z}_{I},\hat{\bar{z}}_{I}\right) & \sim & \mathrm{regular}\,,
\nonumber \\
\bar{T}\left(\bar{z}\right)e^{i\alpha X^{+}}\left(\hat{z}_{I},\hat{\bar{z}}_{I}\right) & \sim & \mathrm{regular}\,,
\end{eqnarray}
for any constant $\alpha$. Therefore $e^{i\alpha X^{+}}\left(\hat{z}_{I},\hat{\bar{z}}_{I}\right)$
is a BRST invariant operator in the conformal gauge bosonic string
theory in noncritical dimensions. 

It is straightforward to define the operator valued supercoordinate
$\hat{\tilde{\mathbf{z}}}_{I}$. We define $\hat{\tilde{\mathbf{z}}}_{I}=\left(\hat{\tilde{z}}_{I},\hat{\tilde{\theta}}_{I}\right)$
to be the operator valued supercoordinate which satisfies
\begin{eqnarray}
\partial\mathcal{X}^{+}\left(\hat{\tilde{\mathbf{z}}}_{I}\right) & = & 0\,,\nonumber \\
\partial D\mathcal{X}^{+}\left(\hat{\tilde{\mathbf{z}}}_{I}\right) & = & 0\,.\label{eq:hatsuperzI}
\end{eqnarray}
Notice that $\hat{\tilde{\mathbf{z}}}_{I}$ is the operator version
of the supercoordinate $\tilde{\mathbf{z}}_{I}$ defined in \cite{Berkovits:1985ji,Berkovits:1987gp,Aoki:1990yn}
which is not superconformal, but it is sufficient for our purpose.
Similarly to the bosonic case, we decompose $\mathcal{X}^{+}\left(\mathbf{z},\bar{\mathbf{z}}\right)$
as
\begin{equation}
\mathcal{X}^{+}\left(\mathbf{z},\bar{\mathbf{z}}\right)=-\frac{i}{2}\left(\rho_{s}\left(\mathbf{z}\right)+\bar{\rho}_{s}\left(\bar{\mathbf{z}}\right)\right)+\delta\mathcal{X}^{+}\left(\mathbf{z},\bar{\mathbf{z}}\right)\,,
\end{equation}
where $\rho_{s},\bar{\rho}_{s}$ are the supersymmetric version of
$\rho,\bar{\rho}$ whose explicit form is given in (\ref{eq:rhos}).
$S_{\alpha}$ in that equation is the one in (\ref{eq:Szego}) or in
(\ref{eq:SalphaL}) according to whether the spin structure of the
fermions is even or odd. Using this decomposition, we get $\hat{\tilde{z}}_{I},\hat{\tilde{\theta}}_{I}$
as expansions around $\tilde{z_{I}}$, $\tilde{\theta}_{I}$ in terms
of the fluctuation $\delta\mathcal{X}^{+}$ as
\begin{eqnarray}
\hat{\tilde{z}}_{I} & = & \tilde{z}_{I}+\sum_{n=1}^{\infty}\delta^{\left(n\right)}\hat{\tilde{z}}_{I}\,,
\nonumber \\
\hat{\tilde{\theta}}_{I} & = & \tilde{\theta}_{I}+\sum_{n=1}^{\infty}\delta^{\left(n\right)}\hat{\tilde{\theta}}_{I}\,,
\end{eqnarray}
assuming $\partial^{2}\rho\left(\tilde{\mathbf{z}}_{I}\right)\ne0$.
For example,
\begin{eqnarray}
\delta^{\left(1\right)}\hat{\tilde{z}}_{I} & = & -\frac{2i}{\partial^{2}\rho}\partial\delta\mathcal{X}^{+}\left(\tilde{\mathbf{z}}_{I}\right)\,,
\nonumber \\
\delta^{\left(1\right)}\hat{\tilde{\theta}}_{I} & = & -\frac{2i}{\partial^{2}\rho}\partial D\delta\mathcal{X}^{+}\left(\tilde{\mathbf{z}}_{I}\right)\,.
\end{eqnarray}
We obtain the OPE's 
\begin{eqnarray}
\begC2{T}\conC{\left(\mathbf{z}\right)}\endC2{\hat{\tilde{z}}_{I}}-\begC2{T}\conC{\left(\mathbf{z}\right)}\endC2{\hat{\tilde{\theta}}_{I}}\hat{\tilde{\theta}}_{I} & \sim & -\frac{\theta-\hat{\tilde{\theta}}_{I}}{\mathbf{z}-\hat{\tilde{\mathbf{z}}}_{I}}-\frac{1}{\left(\mathbf{z}-\hat{\tilde{\mathbf{z}}}_{I}\right)^{2}}\cdot\frac{D\mathcal{X}^{+}}{2\partial^{2}\mathcal{X}^{+}}\left(\hat{\tilde{\mathbf{z}}}_{I}\right)\,,
\nonumber \\
\begC2{T}\conC{\left(\mathbf{z}\right)}\endC2{\hat{\tilde{\theta}}_{I}} & \sim & -\frac{\frac{1}{2}}{\mathbf{z}-\hat{\tilde{\mathbf{z}}}_{I}}-\frac{\theta-\hat{\tilde{\theta}}_{I}}{\left(\mathbf{z}-\hat{\tilde{\mathbf{z}}}_{I}\right)^{3}}\cdot\frac{D\mathcal{X}^{+}}{\partial^{2}\mathcal{X}^{+}}\left(\hat{\tilde{\mathbf{z}}}_{I}\right)
\nonumber \\
& & \quad {}-\frac{1}{\left(\mathbf{z}-\hat{\tilde{\mathbf{z}}}_{I}\right)^{2}}\cdot\frac{\partial^{2}D\mathcal{X}^{+}D\mathcal{X}^{+}}{2\left(\partial^{2}\mathcal{X}^{+}\right)^{2}}\left(\hat{\tilde{\mathbf{z}}}_{I}\right)\,.
\end{eqnarray}
From these OPE's, we get 
\begin{eqnarray}
T\left(\mathbf{z}\right)e^{i\alpha\mathcal{X}^{+}}\left(\hat{\tilde{\mathbf{z}}}_{I},\hat{\bar{\tilde{\mathbf{z}}}}_{I}\right) & \sim & \mathrm{regular}\,,\nonumber \\
\bar{T}\left(\bar{\mathbf{z}}\right)e^{i\alpha\mathcal{X}^{+}}\left(\hat{\tilde{\mathbf{z}}}_{I},\hat{\bar{\tilde{\mathbf{z}}}}_{I}\right) & \sim & \mathrm{regular}\,,\label{eq:superTS}
\end{eqnarray}
for any constant $\alpha$. Therefore $e^{i\alpha\mathcal{X}^{+}}\left(\hat{\tilde{\mathbf{z}}}_{I},\hat{\bar{\tilde{\mathbf{z}}}}_{I}\right)$
is BRST invariant in the conformal gauge superstring theory in noncritical
dimensions. 

Using the operator valued coordinate thus defined, we can define a
BRST invariant operator 
\begin{equation}
e^{-\frac{iQ^{2}}{\alpha_{r}}\mathcal{X}^{+}}\left(\hat{\tilde{\mathbf{z}}}_{I^{(r)}},\hat{\tilde{\bar{\mathbf{z}}}}_{I^{(r)}}\right)\thinspace,
\end{equation}
which can be replaced by
\begin{equation}
e^{-\frac{iQ^{2}}{\alpha_{r}}X^{+}}\left(z_{I^{(r)}},\bar{z}_{I^{(r)}}\right)
\end{equation}
 in evaluating (\ref{eq:DRconformal}), (\ref{eq:ncBRSTodd}) and
(\ref{eq:ncBRST2odd}). This expression can be used instead of the
complicated combination $\oint_{z_{I}^{\left(r\right)}}\frac{dz}{2\pi i}\oint_{\bar{z}_{I}^{\left(r\right)}}\frac{d\bar{z}}{2\pi i}S\left(z,Z_{r}\right)\bar{S}\left(\bar{z},\bar{Z}_{r}\right)$
in \cite{Baba:2009zm,Ishibashi2017b}, which has a similar effect
.

\section{Correlation functions of free fermions\label{sec:Correlation-functions-of}}

In this appendix, we review a few basic facts about the correlation
functions of free fermions on higher genus Riemann surfaces. 

The correlation functions of a free Dirac fermion with spin structure
$\alpha_{\mathrm{L}}$ for left and $\alpha_{\mathrm{R}}$ for
right can be given by \cite{Verlinde:1986kw,D'Hoker:1988ta}
\begin{eqnarray}
 &  & \int\left[d\psi d\bar{\psi}d\bar{\psi}^{\dagger}d\psi^{\dagger}
          \right]_{\hat{g}_{z\bar{z}}}
e^{-S}\psi^{\dagger}(x_{1})\bar{\psi}^{\dagger}(\bar{x}_{1})\cdots\psi^{\dagger}(x_{n})\bar{\psi}^{\dagger}(\bar{x}_{n})\bar{\psi}(\bar{y}_{n})\psi(y_{n})\cdots\bar{\psi}(\bar{y}_{1})\psi(y_{1})\nonumber \\
 &  & \quad=\left(\frac{\det^{\prime}\left(-\hat{g}^{z\bar{z}}\partial_{z}\partial_{\bar{z}}\right)}{\det \mathop{\mathrm{Im}} \Omega\int d^{2}z\sqrt{\hat{g}}}\right)^{-\frac{1}{2}}
\nonumber \\
& & \hphantom{\quad = \quad } 
\times
\vartheta[\alpha_{\mathrm{L}}]\left(\sum_{i=1}^{n}\int_{P_{0}}^{x_{i}}\omega-\sum_{i=1}^{n}\int_{P_{0}}^{y_{i}}\omega\right)\vartheta[\alpha_{\mathrm{R}}]\left(\sum_{i=1}^{n}\int_{P_{0}}^{x_{i}}\omega-\sum_{i=1}^{n}\int_{P_{0}}^{y_{i}}\omega\right)^{\ast}\nonumber \\
 &  & \hphantom{\quad= \quad}
\times\left|\frac{\prod_{i<j}\left[E(x_{i},x_{j})E(y_{j},y_{i})\right]}{\prod_{i,j}E(x_{i},y_{j})}\right|^{2}\,.\label{eq:fermioncorr}
\end{eqnarray}
When both $\alpha_{\mathrm{L}}$ and $\alpha_{\mathrm{R}}$ correspond
to even spin structures, using the formula 
\begin{eqnarray}
 &  & \vartheta\left(\sum_{i=1}^{n}\int_{P_{0}}^{x_{i}}\omega-\sum_{i=1}^{n}\int_{P_{0}}^{y_{i}}\omega-e\right)\vartheta\left(e\right)^{n-1}\frac{\prod_{i<j}\left[E\left(x_{i},x_{j}\right)E\left(y_{j},y_{i}\right)\right]}{\prod_{i,j}E\left(x_{i},y_{j}\right)}\nonumber \\
 &  & \quad=\det\left(\frac{\vartheta\left(\int_{P_{0}}^{x_{i}}\omega-\int_{P_{0}}^{y_{i}}\omega-e\right)}{E\left(x_{i},y_{j}\right)}\right)\,,\label{eq:Fay}
\end{eqnarray}
given in \cite{Fay1973} for the case
\begin{equation}
e_{\nu}=-\left(\Omega\alpha^{\prime}+\alpha^{\prime\prime}\right)_{\nu}\,,
\end{equation}
 it is straightforward to show that (\ref{eq:fermioncorr}) can be
transformed into
\begin{eqnarray}
\lefteqn{
\int\left[d\psi d\bar{\psi}d\bar{\psi}^{\dagger}d\psi^{\dagger}
\right]_{\hat{g}_{z\bar{z}}}
e^{-S}
\psi^{\dagger}(x_{1})\bar{\psi}^{\dagger}(\bar{x}_{1})
  \cdots \psi^{\dagger}(x_{n})\bar{\psi}^{\dagger}(\bar{x}_{n})
\bar{\psi}(\bar{y}_{n})\psi(y_{n})
  \cdots \bar{\psi}(\bar{y}_{1})\psi(y_{1})
}\nonumber \\
 & = &\left(\frac{\det^{\prime}\left(-\hat{g}^{z\bar{z}}\partial_{z}\partial_{\bar{z}}\right)}{\det \mathop{\mathrm{Im}} \Omega\int d^{2}z\sqrt{\hat{g}}}\right)^{-\frac{1}{2}}\vartheta[\alpha_{\mathrm{L}}]\left(0\right)\vartheta[\alpha_{\mathrm{R}}]\left(0\right)^{\ast}\det\left[S_{\alpha_{\mathrm{L}}}\left(x_{i},y_{j}\right)\right]\det\left[S_{\alpha_{\mathrm{R}}}\left(x_{i},y_{j}\right)^{\ast}\right],
~~~~~\label{eq:fermioncorr2}
\end{eqnarray}
 where 
\begin{equation}
S_{\alpha}\left(z,w\right)=\frac{1}{E\left(z,w\right)}\frac{\vartheta\left[\alpha\right]\left(\int_{w}^{z}\omega\right)}{\vartheta\left[\alpha\right]\left(0\right)} 
\label{eq:Szego}
\end{equation}
is the Szego kernel. The expression (\ref{eq:fermioncorr2}) implies
that the partition function is given by 
\begin{equation}
\left(Z^{\psi}[g_{z\bar{z}}]\right)^{2}=\left(\frac{\det^{\prime}\left(-g^{z\bar{z}}\partial_{z}\partial_{\bar{z}}\right)}{\det \mathop{\mathrm{Im}} \Omega\int d^{2}z\sqrt{g}}\right)^{-\frac{1}{2}}\vartheta[\alpha_{\mathrm{L}}]\left(0\right)\vartheta[\alpha_{\mathrm{R}}]\left(0\right)^{\ast}\,,\label{eq:Zpsi}
\end{equation}
and the propagators of the fermions are 
\begin{eqnarray}
\begC1{\psi^{\dagger}}\conC{(x)}\endC1{\psi}(y) & = & S_{\alpha_{\mathrm{L}}}\left(x,y\right)\,,
\nonumber \\
\begC1{\bar{\psi}^{\dagger}}\conC{(\bar{x})}\endC1{\bar{\psi}}(\bar{y}) & = & S_{\alpha_{\mathrm{R}}}\left(x,y\right)^{\ast}\,.
\end{eqnarray}
When the spin structures are not even, we need to take care of the
fermion zero modes. For example, let us consider the case where $\alpha_{\mathrm{L}}$
corresponds to an odd spin structure and $\alpha_{\mathrm{R}}$ corresponds
to an even one. In this case, using the formula (\ref{eq:Fay}) for
\begin{equation}
e_{\nu}=-\left(\Omega\alpha_{\mathrm{L}}^{\prime}+\alpha_{\mathrm{L}}^{\prime\prime}\right)_{\nu}-\int_{q}^{p}\omega_{\nu}\,,
\end{equation}
in the limit $q\to p$, we get \cite{Fay1973}
\begin{eqnarray}
\lefteqn{
\vartheta[\alpha_{\mathrm{L}}]
   \left( \sum_{i=1}^{n}\int_{P_{0}}^{x_{i}}\omega
          -\sum_{i=1}^{n}\int_{P_{0}}^{y_{i}}\omega\right)
\frac{\prod_{i<j}\left[E(x_{i},x_{j})E(y_{j},y_{i})\right]}
     {\prod_{i,j}E(x_{i},y_{j})}
}\nonumber \\
 &= &\int d\psi_{0}d\psi_{0}^{\dagger}
\, \det \left[\psi_{0}^{\dagger}h_{\alpha_{\mathrm{L}}}(x_{i})
               \psi_{0}h_{\alpha_{\mathrm{L}}}(y_{j})
             + \frac{1}{E\left(x_{i},y_{j}\right)}
                 \frac{\sum_{\nu}\partial_{\nu}
                       \vartheta\left[\alpha_{\mathrm{L}}\right]
                          \left(\int_{y_{j}}^{x_{i}}\omega\right)
                       \omega_{\nu}(p)}
                      {\sum_{\nu}\partial_{\nu}
                        \vartheta\left[\alpha_{\mathrm{L}}\right]
                          \left( 0 \right)
                        \omega_{\nu}(p)}
          \right].~~~~
\label{eq:Fay2}
\end{eqnarray}
Here 
\begin{equation}
h_{\alpha_{\mathrm{L}}}(z)=\sqrt{\sum_{\nu}\partial_{\nu}\vartheta\left[\alpha_{\mathrm{L}}\right]\left(0\right)\omega_{\nu}(z)}
\label{eq:halphaL}
\end{equation}
gives the zero mode of the fermion. Substituting (\ref{eq:Fay2})
into the right hand side of (\ref{eq:fermioncorr2}), we obtain
\begin{eqnarray}
 &  & \int\left[d\psi d\bar{\psi}d\bar{\psi}^{\dagger}d\psi^{\dagger}
          \right]_{\hat{g}_{z\bar{z}}}
e^{-S}\psi^{\dagger}(x_{1})\bar{\psi}^{\dagger}(\bar{x}_{1})\cdots\psi^{\dagger}(x_{n})\bar{\psi}^{\dagger}(\bar{x}_{n})\bar{\psi}(\bar{y}_{n})\psi(y_{n})\cdots\bar{\psi}(\bar{y}_{1})\psi(y_{1})\nonumber \\
 &  & \quad=\left(\frac{\det^{\prime}\left(-\hat{g}^{z\bar{z}}\partial_{z}\partial_{\bar{z}}\right)}{\det\mathop{\mathrm{Im}} \Omega\int d^{2}z\sqrt{\hat{g}}}\right)^{-\frac{1}{2}}
\int d\psi_{0}d\psi_{0}^{\dagger}\det\left[S_{\alpha_{\mathrm{L}}}\left(x_{i},y_{j}\right)\right]\vartheta[\alpha_{\mathrm{R}}]\left(0\right)^{\ast}\det\left[S_{\alpha_{\mathrm{R}}}\left(x_{i},y_{j}\right)^{\ast}\right],
\nonumber \\
& & \label{eq:freeodd}
\end{eqnarray}
where 
\begin{equation}
S_{\alpha_{\mathrm{L}}}(x,y)=\psi_{0}^{\dagger}h_{\alpha_{\mathrm{L}}}(x)\psi_{0}h_{\alpha_{\mathrm{L}}}(y)+\frac{1}{E\left(x,y\right)}\frac{\sum_{\nu}\partial_{\nu}\vartheta\left[\alpha_{\mathrm{L}}\right]\left(\int_{y}^{x}\omega\right)\omega_{\nu}(p)}{\sum_{\nu}\partial_{\nu}\vartheta\left[\alpha_{\mathrm{L}}\right]\left(0\right)\omega_{\nu}(p)}\,.\label{eq:SalphaL}
\end{equation}
$S_{\alpha_{\mathrm{L}}}(x,y)$ can be identified with the propagator
of the left-moving fermions and it involves the zero mode variables
$\psi_{0}^{\dagger},\psi_{0}$. $\psi_{0}^{\dagger}$ and $\psi_{0}$
should be integrated over after all the contractions are performed.
The other cases where $\alpha_{\mathrm{R}}$ corresponds to an odd
spin structure can be dealt with in the same way.

\section{Dimensional regularization for odd spin structure}

In this appendix, we explain the details of how dimensional regularization
works in the case of odd spin structure.

\subsection{Supersymmetric $X^{\pm}$ CFT\label{subsec:Supersymmetric--CFT}}

In order to get the expression of the amplitudes in the conformal
gauge, we need to calculate the correlation functions of the supersymmetric
$X^{\pm}$ CFT on the surface with odd spin structure. 

The action of the supersymmetric $X^{\pm}$ CFT is given in the form
\begin{equation}
S_{\mathrm{super}}^{\pm}\left[\hat{g}_{z\bar{z}},\mathcal{X}^{\pm}\right]=S_{\mathrm{\mathrm{free}}}\left[\hat{g}_{z\bar{z}},\mathcal{X}^{\pm}\right]+\frac{d-10}{8}\Gamma_{\mathrm{super}}\left[\hat{g}_{z\bar{z}},\,2i\mathcal{X}^{+}\right]\,,
\end{equation}
where $S_{\mathrm{\mathrm{free}}}\left[\hat{g}_{z\bar{z}},\mathcal{X}^{\pm}\right]$
denotes the free action of $\mathcal{X}^{\pm}$. When the spin structures
are both even, the correlation functions of the supersymmetric $X^{\pm}$
CFT are evaluated as \cite{Ishibashi2017b}
\begin{eqnarray}
 &  & \int\left[d\mathcal{X}^{+}d\mathcal{X}^{-}\right]_{\hat{g}_{z\bar{z}}}
  e^{-S_{\mathrm{super}}^{\pm}\left[\hat{g}_{z\bar{z}},
                           \mathcal{X}^{\pm}\right]}
 \prod_{r=1}^{N} e^{-ip_{r}^{+}\mathcal{X}^{-}}
                     (\mathbf{Z}_{r},\bar{\mathbf{Z}}_{r})
 \prod_{u=1}^{M}e^{-ip_{u}^{-}\mathcal{X}^{+}}
                    (\mathbf{w}_{u},\bar{\mathbf{w}}_{u})
\nonumber \\
 &  & \quad=\int \left[d\mathcal{X}^{+}d\mathcal{X}^{-}
                 \right]_{\hat{g}_{z\bar{z}}}e^{-S_{\mathrm{\mathrm{free}}}
        \left[\hat{g}_{z\bar{z}},\mathcal{X}^{\pm}\right]}
\prod_{r=1}^{N}e^{-ip_{r}^{+}\mathcal{X}^{-}}(\mathbf{Z}_{r},\bar{\mathbf{Z}}_{r}) \nonumber \\
& & \hphantom{
     \quad=\int\left[d\mathcal{X}^{+}d\mathcal{X}^{-}
                \right]_{\hat{g}_{z\bar{z}}}e^{-S_{\mathrm{\mathrm{free}}}}
    }
\qquad 
\times e^{-\frac{d-10}{8}\Gamma_{\mathrm{super}}
             \left[\hat{g}_{z\bar{z}},\,2i\mathcal{X}^{+}\right]}
\prod_{u=1}^{M}e^{-ip_{u}^{-}\mathcal{X}^{+}}
                 (\mathbf{w}_{u},\bar{\mathbf{w}}_{u})
\nonumber \\
 &  & \quad=(2\pi)^{2}\delta\left(\sum_{u}p_{u}^{-}\right)\delta\left(\sum_{r}p_{r}^{+}\right)
\left(Z_{\mathrm{super}}^{\mathcal{X}}[\hat{g}_{z\bar{z}}]\right)^{2}
\nonumber \\
& & \hspace{8em} \times
e^{-\frac{d-10}{8}
    \Gamma_{\mathrm{super}}
      \left[\hat{g}_{z\bar{z}},\,\rho_{s}+\bar{\rho}_{s}\right]} \, 
\prod_{u}
  e^{-p_{u}^{-} \frac{\rho_{s}+\bar{\rho}_{s}}{2}}
      (\mathbf{w}_{u},\bar{\mathbf{w}}_{u})~.
\label{eq:superXpmCFTcorr}
\end{eqnarray}
Regarding the second and third lines as a correlation function of
\begin{equation}
e^{-\frac{d-10}{8}
    \Gamma_{\mathrm{super}}\left[\hat{g}_{z\bar{z}},\,2i\mathcal{X}^{+}\right]}
\prod_{u=1}^{M}
   e^{-ip_{u}^{-}\mathcal{X}^{+}} (\mathbf{w}_{u},\bar{\mathbf{w}}_{u})
\end{equation}
of the free theory with the source term
\begin{equation}
\prod_{r=1}^{N}e^{-ip_{r}^{+}\mathcal{X}^{-}}(\mathbf{Z}_{r},\bar{\mathbf{Z}}_{r})\,,
\end{equation}
we can calculate it by replacing the $\mathcal{X}^{+}\left(\mathbf{z},\bar{\mathbf{z}}\right)$
by its expectation value $-\frac{i}{2}\left(\rho_{s}\left(\mathbf{z}\right)+\bar{\rho}_{s}\left(\bar{\mathbf{z}}\right)\right)$
and derive the fourth line. Here $\rho_{s},\bar{\rho}_{s}$ are the
supersymmetric version of $\rho,\bar{\rho}$ and expressed as
\begin{eqnarray}
\rho_{s}\left(\mathbf{z}\right) & = & \rho\left(z\right)-\theta\sum_{r}\alpha_{r}\Theta_{r}S_{\alpha_{\mathrm{L}}}\left(z,Z_{r}\right)\,,\nonumber \\
\bar{\rho}_{s}(\bar{\mathbf{z}}) & = & \bar{\rho}\left(\bar{z}\right)-\bar{\theta}\sum_{r}\alpha_{r}\bar{\Theta}_{r}S_{\alpha_{\mathrm{R}}}\left(\bar{z},\bar{Z}_{r}\right)\,,\label{eq:rhos}
\end{eqnarray}
where $S_{\alpha_{\mathrm{L}}}$ and $S_{\alpha_{\mathrm{R}}}$ are
taken to be the Szego kernel (\ref{eq:Szego}). The partition function
$\left(Z_{\mathrm{super}}^{\mathcal{X}}[\hat{g}_{z\bar{z}}]\right)^{2}$
is described by using $Z_{X^{\pm}}\left[\hat{g}_{z\bar{z}} \right]$
and $Z_{\psi^{\pm}} \left[ \hat{g}_{z\bar{z}} \right]$ 
in (\ref{eq:psipm}) as
\begin{equation}
\left(Z_{\mathrm{super}}^{\mathcal{X}}[\hat{g}_{z\bar{z}}]\right)^{2}
=
Z_{X^{\pm}} \left[ \hat{g}_{z\bar{z}} \right]
Z_{\psi^{\pm}} \left[ \hat{g}_{z\bar{z}} \right] \,.
\end{equation}
The explicit form of $e^{-\frac{d-10}{8}\Gamma_{\mathrm{super}}\left[\hat{g}_{z\bar{z}},\,\rho_{s}+\bar{\rho}_{s}\right]}$
can be found in \cite{Ishibashi2017b}.

In the case where $\alpha_{\mathrm{L}}$ corresponds to an odd spin
structure, we can proceed in the same way. Since the correlation functions
of the free fermions are given in (\ref{eq:freeodd}) as an integral
over the zero modes $\psi_{0}^{\pm}$, we obtain
\begin{eqnarray}
\lefteqn{
\int\left[d\mathcal{X}^{+}d\mathcal{X}^{-}\right]_{\hat{g}_{z\bar{z}}}
 e^{-S_{\mathrm{super}}^{\pm}\left[\hat{g}_{z\bar{z}}\right]}
 \prod_{r=1}^{N} e^{-ip_{r}^{+}\mathcal{X}^{-}}
                   (\mathbf{Z}_{r},\bar{\mathbf{Z}}_{r})
\prod_{u=1}^{M} e^{-ip_{u}^{-}\mathcal{X}^{+}}
                  (\mathbf{w}_{u},\bar{\mathbf{w}}_{u})}
\nonumber \\
 & =& (2\pi)^{2}
   \delta\left(\sum_{u}p_{u}^{-}\right)
   \delta\left(\sum_{r}p_{r}^{+}\right)
Z_{X^{\pm}} \left[\hat{g}_{z\bar{z}} \right]
\left(\frac{\det^{\prime}\left(-\hat{g}^{z\bar{z}}
                              \partial_{z}\partial_{\bar{z}}\right)}
           {\det\mathop{\mathrm{Im}}\Omega
              \int d^{2}z\sqrt{\hat{g}}}\right)^{-\frac{1}{2}}
\vartheta[\alpha_{\mathrm{R}}]\left(0\right)^{\ast}
\nonumber \\
 &  & 
\hspace{5em}
\times\int d\psi_{0}^{+}d\psi_{0}^{-}
   \prod_{u} e^{-p_{u}^{-} \frac{\rho_{s}+\bar{\rho}_{s}}{2}}
               (\mathbf{w}_{u},\bar{\mathbf{w}}_{u})
\,e^{-\frac{d-10}{8}
      \Gamma_{\mathrm{super}}
         \left[\hat{g}_{z\bar{z}},\,\rho_{s}+\bar{\rho}_{s}\right]}
\,,\label{eq:superXpmCFTcorrodd}
\end{eqnarray}
where $\rho_{s},\bar{\rho}_{s}$ in this formula are (\ref{eq:rhos})
with $S_{\alpha_{\mathrm{L}}}$ given in (\ref{eq:SalphaL}) and $S_{\alpha_{\mathrm{R}}}$
taken to be the Szego kernel (\ref{eq:Szego}).

With the correlation function (\ref{eq:superXpmCFTcorrodd}), it is
straightforward to check the following properties of the energy-momentum
tensor:
\begin{itemize}
\item $T^{\mathcal{X}^{\pm}}\left(\mathbf{z}\right)$ is regular at $z=z_{I}$.
\item $T^{\mathcal{X}^{\pm}}\left(\mathbf{z}\right)$ satisfies
\begin{eqnarray}
T^{\mathcal{X}^{\pm}}\left(\mathbf{z}\right)e^{-ip_{r}^{+}\mathcal{X}^{-}-ip_{r}^{-}\mathcal{X}^{+}}(\mathbf{Z}_{r},\bar{\mathbf{Z}}_{r})
& \sim & \frac{\theta-\Theta_{r}}{\left(z-Z_{r}\right)^{2}}\left(-p_{r}^{+}p_{r}^{-}\right)e^{-ip_{r}^{+}\mathcal{X}^{-}-ip_{r}^{-}\mathcal{X}^{+}}(\mathbf{Z}_{r},\bar{\mathbf{Z}}_{r})
\nonumber \\
 &  & \ \ {}+\frac{1}{\mathbf{z}-\mathbf{Z}_{r}}
   \frac{1}{2} De^{-ip_{r}^{+}\mathcal{X}^{-}-ip_{r}^{-}\mathcal{X}^{+}}
                    (\mathbf{Z}_{r},\bar{\mathbf{Z}}_{r})
\nonumber \\
 &  & \ \  {}+\frac{\theta-\Theta_{r}}{z-Z_{r}}
   \partial e^{-ip_{r}^{+}\mathcal{X}^{-}-ip_{r}^{-}\mathcal{X}^{+}}
                (\mathbf{Z}_{r},\bar{\mathbf{Z}}_{r})\,.
\end{eqnarray}
\item The OPE between $\mathcal{X}^{-}$'s is given by 
\begin{eqnarray}
 &  & D\mathcal{X}^{-}\left(\mathbf{z}\right)D\mathcal{X}^{-}\left(\mathbf{z}'\right) 
\nonumber \\
 &  & \quad\sim-\frac{d-10}{4}
\nonumber \\
& & \qquad \  \times  
DD^{\prime}\left[\frac{\theta-\theta'}{\left(\mathbf{z}-\mathbf{z}'\right)^{3}}\frac{3D\mathcal{X}^{+}}{\left(\partial\mathcal{X}^{+}\right)^{3}}\left(\mathbf{z}'\right)\right.
\nonumber \\
 &  & \hphantom{\qquad   \times DD^{\prime}}
{}+\frac{1}{\left(\mathbf{z}-\mathbf{z}'\right)^{2}}\left(\frac{1}{2\left(\partial\mathcal{X}^{+}\right)^{2}}+\frac{4\partial D\mathcal{X}^{+}D\mathcal{X}^{+}}{\left(\partial\mathcal{X}^{+}\right)^{4}}\right)\left(\mathbf{z}'\right)
\nonumber \\
 &  & \hphantom{\qquad \times DD^{\prime}}
{}+\frac{\theta-\theta'}{\left(\mathbf{z}-\mathbf{z}'\right)^{2}}\left(-\frac{\partial D\mathcal{X}^{+}}{\left(\partial\mathcal{X}^{+}\right)^{3}}-\frac{5\partial^{2}\mathcal{X}^{+}D\mathcal{X}^{+}}{2\left(\partial\mathcal{X}^{+}\right)^{4}}\right)\left(\mathbf{z}'\right)
\nonumber \\
 &  & \hphantom{\qquad  \times DD^{\prime}}
{}+\frac{1}{\mathbf{z}-\mathbf{z}'}\left(-\frac{\partial^{2}\mathcal{X}^{+}}{2\left(\partial\mathcal{X}^{+}\right)^{3}}+\frac{2\partial^{2}D\mathcal{X}^{+}D\mathcal{X}^{+}}{\left(\partial\mathcal{X}^{+}\right)^{4}}-\frac{8\partial^{2}\mathcal{X}^{+}\partial D\mathcal{X}^{+}D\mathcal{X}^{+}}{\left(\partial\mathcal{X}^{+}\right)^{5}}\right)\left(\mathbf{z}'\right)
\nonumber \\
 &  & \hphantom{\qquad  \times DD^{\prime}}
{}+\frac{\theta-\theta'}{\mathbf{z}-\mathbf{z}'}\left(-\frac{\partial^{2}D\mathcal{X}^{+}}{2\left(\partial\mathcal{X}^{+}\right)^{3}}+\frac{3\partial^{2}\mathcal{X}^{+}\partial D\mathcal{X}^{+}}{2\left(\partial\mathcal{X}^{+}\right)^{4}}-\frac{\partial^{3}\mathcal{X}^{+}D\mathcal{X}^{+}}{2\left(\partial\mathcal{X}^{+}\right)^{4}}\right.
\nonumber \\
 &  & \hphantom{\qquad \times DD^{\prime}+\frac{\theta-\theta'}{\mathbf{z}-\mathbf{z}'}}\left.\left.{}+\frac{\left(\partial^{2}\mathcal{X}^{+}\right)^{2}D\mathcal{X}^{+}}{\left(\partial\mathcal{X}^{+}\right)^{5}}-\frac{\partial^{2}D\mathcal{X}^{+}\partial D\mathcal{X}^{+}D\mathcal{X}^{+}}{\left(\partial\mathcal{X}^{+}\right)^{5}}\right)\left(\mathbf{z}'\right)\right],
\end{eqnarray}
and we can deduce that the energy momentum tensor $T^{\mathcal{X}^{\pm}}\left(\mathbf{z}\right)$
satisfies the OPE
\begin{eqnarray}
 &  & T^{\mathcal{X}^{\pm}}\left(\mathbf{z}\right)T^{\mathcal{X}^{\pm}}\left(\mathbf{z}^{\prime}\right)
\nonumber \\
 &  & \quad\sim\frac{12-d}{4\left(\mathbf{z}-\mathbf{z}^{\prime}\right)^{3}}+\frac{\theta-\theta^{\prime}}{\left(\mathbf{z}-\mathbf{z}^{\prime}\right)^{2}}\frac{3}{2}T^{\mathcal{X}^{\pm}}\left(\mathbf{z}^{\prime}\right)+\frac{1}{\mathbf{z}-\mathbf{z}^{\prime}}\frac{1}{2}DT^{\mathcal{X}^{\pm}}\left(\mathbf{z}^{\prime}\right)+\frac{\theta-\theta^{\prime}}{\mathbf{z}-\mathbf{z}^{\prime}}\partial T^{\mathcal{X}^{\pm}}\left(\mathbf{z}^{\prime}\right),
\nonumber \\
\end{eqnarray}
which corresponds to the super Virasoro algebra with the central charge
$\hat{c}=12-d$. It follows that combined with the transverse variables
$X^{i}\left(\mathbf{z},\mathbf{\bar{z}}\right)$, the total central
charge of the system becomes $\hat{c}=10$. This implies that with
the ghost superfields $B\left(\mathbf{z}\right)$ and $C\left(\mathbf{z}\right)$
defined as
\begin{equation}
B(\mathbf{z})=\beta(z)+\theta b(z)\;,
\qquad C(\mathbf{z})=c(z)+\theta\gamma(z)\;,
\end{equation}
it is possible to construct a nilpotent BRST charge
\begin{equation}
Q_{\mathrm{B}}=\oint\frac{d\mathbf{z}}{2\pi i}\left[-C\left(T^{\mathcal{X}^{\pm}}-\frac{1}{2}D\mathcal{X}^{i}\partial\mathcal{X}^{i}\right)+\left(C\partial C-\frac{1}{4}\left(DC\right)^{2}\right)B\right].\label{eq:BRSTcharge}
\end{equation}
\end{itemize}
These properties can be proved in the same way as in the even spin
structure case, because we need only the behaviors of the fermion
propagators around the singularities to do so.

In the same way as in the even spin structure case \cite{Ishibashi2017b},
we can derive from (\ref{eq:superXpmCFTcorrodd}), 
\begin{eqnarray}
\lefteqn{
\int\left[d\mathcal{X}^{+}d\mathcal{X}^{-}\right]_{\hat{g}_{z\bar{z}}}e^{-S_{\mathrm{super}}^{\pm}\left[\hat{g}_{z\bar{z}}\right]}\prod_{r=1}^{N}e^{-ip_{r}^{+}X^{-}}(Z_{r},\bar{Z}_{r})\prod_{s=1}^{M}e^{-ip_{s}^{-}X^{+}}(w_{s},\bar{w}_{s})
} \nonumber \\
 &  & \hphantom{
               \int\left[d\mathcal{X}^{+}d\mathcal{X}^{-}
                    \right]_{\hat{g}_{z\bar{z}}}
                e^{-S_{\mathrm{super}}^{\pm}
               } }
\times
  \psi^{+}\left(u_{1}\right)\cdots\psi^{+}\left(u_{n}\right)
  \psi^{-}\left(v_{1}\right)\cdots\psi^{-}\left(v_{m}\right)
\nonumber \\
 &  & \hphantom{
            \int\left[d\mathcal{X}^{+}d\mathcal{X}^{-}
                \right]_{\hat{g}_{z\bar{z}}}
            e^{-S_{\mathrm{super}}^{\pm}
              } }
\times
  \bar{\psi}^{+}\left(\tilde{u}_{1}\right)\cdots
              \bar{\psi}^{+}\left(\tilde{u}_{n}\right)
  \bar{\psi}^{-}\left(\tilde{v}_{1}\right)\cdots
              \bar{\psi}^{-}\left(\tilde{v}_{m}\right)
\nonumber \\
 &=& (2\pi)^{2} \delta\left(\sum_{s}p_{s}^{-}\right)
                \delta\left(\sum_{r}p_{r}^{+}\right)
Z_{X^{\pm}} \left[ \hat{g}_{z\bar{z}} \right]
\prod_{s}e^{-p_{s}^{-}  \frac{1}{2}
                  \left(\mathbf{\rho}+\bar{\mathbf{\rho}}\right)}
            (w_{s},\bar{w}_{s})
e^{-\frac{d-10}{16} \Gamma \left[\sigma;\hat{g}_{z\bar{z}} \right]}
\nonumber \\
 &  & 
\qquad 
\times \int \left[d\psi^{+}d\psi^{-}d\bar{\psi}^{+}d\bar{\psi}^{-}
            \right]_{\hat{g}_{z\bar{z}}}
   e^{\frac{1}{\pi}\int d^{2}z\left(\psi^{-}\bar{\partial}\psi^{+}+\bar{\psi}^{-}\partial\bar{\psi}^{+}\right)-S_{\mathrm{int}}}
\nonumber \\
 &  & \hphantom{= \times\int
                \left[d\psi^{+}d\psi^{-}d\bar{\psi}^{+}d\bar{\psi}^{-}
                \right] \quad }
\times
    \psi^{+}\left(u_{1}\right)\cdots\psi^{+}\left(u_{n}\right)
    \psi^{-}\left(v_{1}\right)\cdots\psi^{-}\left(v_{m}\right)
\nonumber \\
 &  & \hphantom{=\times  \int
                 \left[d\psi^{+}d\psi^{-}d\bar{\psi}^{+}d\bar{\psi}^{-}
                 \right]\quad}
\times\bar{\psi}^{+}\left(\tilde{u}_{1}\right)\cdots
          \bar{\psi}^{+}\left(\tilde{u}_{n}\right)
      \bar{\psi}^{-}\left(\tilde{v}_{1}\right)\cdots
          \bar{\psi}^{-}\left(\tilde{v}_{m}\right)\,,
\label{eq:fermioniccorrodd}
\end{eqnarray}
where 
\begin{eqnarray}
S_{\mathrm{int}} & = & \frac{d-10}{8}\left[-\sum_{r}\frac{2}{\alpha_{r}}\frac{\partial\psi^{+}\psi^{+}}{\partial^{2}\rho}\left(z_{I^{\left(r\right)}}\right)\vphantom{\left\{ \left(\frac{5}{3}\frac{\partial^{4}\rho_{b}}{\left(\partial^{2}\rho_{b}\right)^{3}}-3\frac{\left(\partial^{3}\rho_{b}\right)^{2}}{\left(\partial^{2}\rho_{b}\right)^{4}}\right)\partial\psi^{+}\psi^{+}-\frac{8}{3}\frac{\partial^{3}\psi^{+}\psi^{+}}{\left(\partial^{2}\rho_{b}\right)^{2}}+\frac{4\partial^{3}\rho_{b}}{\left(\partial^{2}\rho_{b}\right)^{3}}\partial^{2}\psi^{+}\psi^{+}+\frac{4}{3}\frac{\partial^{3}\psi^{+}\partial^{2}\psi^{+}\partial\psi^{+}\psi^{+}}{\left(\partial^{2}\rho_{b}\right)^{4}}\right\} }\right.
\nonumber \\
 &  & \hphantom{\frac{d-10}{8}\quad}
{}+\sum_{I}\left\{ \left(\frac{5}{3}\frac{\partial^{4}\rho}{\left(\partial^{2}\rho\right)^{3}}-3\frac{\left(\partial^{3}\rho\right)^{2}}{\left(\partial^{2}\rho\right)^{4}}\right)\partial\psi^{+}\psi^{+}-\frac{8}{3}\frac{\partial^{3}\psi^{+}\psi^{+}}{\left(\partial^{2}\rho\right)^{2}}\right.
\nonumber \\
 &  & \hphantom{\frac{d-10}{8}\quad+\sum_{I}
     \left\{ \right\}
  }
\left. {}+\frac{4\partial^{3}\rho}{\left(\partial^{2}\rho\right)^{3}}\partial^{2}\psi^{+}\psi^{+}+\frac{4}{3}\frac{\partial^{3}\psi^{+}\partial^{2}\psi^{+}\partial\psi^{+}\psi^{+}}{\left(\partial^{2}\rho\right)^{4}}\right\} \left(z_{I}\right)
\nonumber \\
 &  & \hphantom{\frac{d-10}{8}\quad}\left.+\mathrm{c.c.}\vphantom{\left\{ \left(\frac{5}{3}\frac{\partial^{4}\rho_{b}}{\left(\partial^{2}\rho_{b}\right)^{3}}-3\frac{\left(\partial^{3}\rho_{b}\right)^{2}}{\left(\partial^{2}\rho_{b}\right)^{4}}\right)\partial\psi^{+}\psi^{+}-\frac{8}{3}\frac{\partial^{3}\psi^{+}\psi^{+}}{\left(\partial^{2}\rho_{b}\right)^{2}}+\frac{4\partial^{3}\rho_{b}}{\left(\partial^{2}\rho_{b}\right)^{3}}\partial^{2}\psi^{+}\psi^{+}+\frac{4}{3}\frac{\partial^{3}\psi^{+}\partial^{2}\psi^{+}\partial\psi^{+}\psi^{+}}{\left(\partial^{2}\rho_{b}\right)^{4}}\right\} }\right]\,.
\end{eqnarray}
The path integral over $\psi^{\pm},\bar{\psi}^{\pm}$ can be computed
by treating $S_{\mathrm{int}}$ perturbatively. Since $S_{\mathrm{int}}$
involves only $\psi^{+}$, the perturbation series terminates at a
finite order. 

\subsection{A proof of equality of (\ref{eq:ncBRSTodd}) and (\ref{eq:ncBRST2odd})\label{subsec:A-proof-of}}

In this appendix, we show that (\ref{eq:ncBRSTodd}) is equal to (\ref{eq:ncBRST2odd}). 
In the case $Q=0$, this equality implies that  (\ref{eq:oddBRST}) is equal to (\ref{eq:oddBRST2}). 
Proving this can be done by using 
a fermionic charge\footnote{This fermionic charge 
was used in \cite{Ishibashi:2010nq}.}
\begin{equation}
\hat{Q}^{\prime}\equiv\oint\frac{dz}{2\pi i}\left[-\frac{b}{4\partial\rho}\left(iX_{\mathrm{L}}^{+}-\frac{1}{2}\rho\right)\left(z\right)+\frac{\beta}{2\partial\rho}\psi^{+}\left(z\right)\right]\,,
\end{equation}
and its antiholomorphic counterpart $\hat{\bar{Q}}^{\prime}$. Here
$\left(iX_{\mathrm{L}}^{+}-\frac{1}{2}\rho\right)\left(z\right)$
is defined as
\begin{equation}
\left(iX_{\mathrm{L}}^{+}-\frac{1}{2}\rho\right)\left(z\right)\equiv\int_{w_{0}}^{z}dz^{\prime}\left(i\partial X^{+}-\frac{1}{2}\partial\rho\right)\left(z^{\prime}\right)\,,
\end{equation}
with a generic point $w_{0}$ on the surface. $\left(iX_{\mathrm{L}}^{+}-\frac{1}{2}\rho\right)\left(z\right)$
thus defined is single valued on the surface in the correlation functions
we consider here because $-\frac{i}{2}\rho$ coincides with the expectation
value of $X_{\mathrm{L}}^{+}$ in the presence of 
the sources $e^{-ip_{r}^{+}X^{-}}$.
In order to use $\hat{Q}^{\prime}$, we need to rewrite the ghost
part of the correlation function. Inserting
\begin{equation}
1=\left|\oint_{w_{0}}\frac{dz}{2\pi i}\frac{b}{\partial\rho}\left(z\right)\partial\rho c\left(w_{0}\right)\right|^{2}
\end{equation}
into (\ref{eq:oddBRST}) and deforming the contours of the antighost
insertions, (\ref{eq:oddBRST}) is transformed into 
\begin{eqnarray}
 &  & \int\left[dX^{\mu}d\psi^{\mu}d\bar{\psi}^{\mu}dbd\bar{b}dcd\bar{c}d\beta d\bar{\beta}d\gamma d\bar{\gamma}\right]_{g_{z\bar{z}}^{\mathrm{A}}}e^{-S^{\mathrm{tot}}}\nonumber \\
 &  & \qquad\times\partial\rho c\left(w_{0}\right)\bar{\partial}\bar{\rho}\bar{c}\left(\bar{w}_{0}\right)\nonumber \\
 &  & \qquad\times\prod_{j=1}^{g}\left[\left(\oint_{\alpha_{j}}\frac{dz}{\partial\rho}b_{zz}+\oint_{\alpha_{j}}\frac{d\bar{z}}{\bar{\partial}\bar{\rho}}b_{\bar{z}\bar{z}}\right)\left(\oint_{\beta_{j}}\frac{dz}{\partial\rho}b_{zz}+\oint_{\beta_{j}}\frac{d\bar{z}}{\bar{\partial}\bar{\rho}}b_{\bar{z}\bar{z}}\right)\right]\nonumber \\
 &  & \qquad\times\prod_{I}\left[\oint_{z_{I}}\frac{dz}{2\pi i}\frac{b}{\partial\rho}\left(z\right)X\left(z_{I}\right)\oint_{\bar{z}_{I}}\frac{d\bar{z}}{2\pi i}\frac{\bar{b}}{\partial\bar{\rho}}\left(\bar{z}\right)\bar{X}\left(\bar{z}_{I}\right)\right]\prod_{r}e^{-\frac{iQ^{2}}{\alpha_{r}}\mathcal{X}^{+}}\left(\hat{\tilde{\mathbf{z}}}_{I^{\left(r\right)}},\hat{\tilde{\bar{\mathbf{z}}}}_{I^{\left(r\right)}}\right)\nonumber \\
 &  & \qquad\times V_{1}^{\left(-2,-1\right)}\left(Z_{1},\bar{Z}_{1}\right)V_{2}^{\left(0,-1\right)}\left(Z_{2},\bar{Z}_{2}\right)\prod_{r=3}^{N}\left[V_{r}^{\left(-1,-1\right)}(Z_{r},\bar{Z}_{r})\right]\,.\label{eq:oddBRST3}
\end{eqnarray}
Here $\alpha_{j}$ and $\beta_{j}$ are chosen so that they form a
canonical basis of the first homology group of the Riemann surface.
Using $\hat{Q}^{\prime}$, the operators inserted at $z=z_{I}$ can
be expressed as 
\begin{eqnarray}
 &  & \oint_{z_{I}}\frac{dz}{2\pi i}\frac{b}{\partial\rho}\left(z\right)X\left(z_{I}\right)\sideset{}{^{\prime}}\prod_{r}e^{-\frac{iQ^{2}}{\alpha_{r}}\mathcal{X}^{+}}\left(\hat{\tilde{\mathbf{z}}}_{I^{\left(r\right)}},\hat{\tilde{\bar{\mathbf{z}}}}_{I^{\left(r\right)}}\right)\nonumber \\
 &  & \quad=-\oint_{z_{I}}\frac{dz}{2\pi i}\frac{b}{\partial\rho}\left(z\right)e^{\phi}T_{\mathrm{F}}^{\mathrm{LC}}\left(z_{I}\right)\sideset{}{^{\prime}}\prod_{r}e^{-\frac{iQ^{2}}{\alpha_{r}} \mathcal{X}^{+}}\left(\hat{\tilde{\mathbf{z}}}_{I^{\left(r\right)}},\hat{\tilde{\bar{\mathbf{z}}}}_{I^{\left(r\right)}}\right)\nonumber \\
 &  & \hphantom{\quad=}-\left\{ \hat{Q}^{\prime},\oint_{z_{I}}\frac{dz}{2\pi i}\frac{b}{\partial\rho}\left(z\right)\oint_{z_{I}}\frac{dw}{2\pi i}\frac{A\left(w\right)}{w-z_{I}}e^{\phi}\left(z_{I}\right)\sideset{}{^{\prime}}\prod_{r}e^{-\frac{iQ^{2}}{\alpha_{r}}\mathcal{X}^{+}}\left(\hat{\tilde{\mathbf{z}}}_{I^{\left(r\right)}},\hat{\tilde{\bar{\mathbf{z}}}}_{I^{\left(r\right)}}\right)\right\} \nonumber \\
 &  & \hphantom{\quad=}+\frac{1}{4}\oint_{z_{I}}\frac{dz}{2\pi i}\frac{b}{\partial\rho}\left(z\right)\oint_{z_{I}}\frac{dw}{2\pi i}\frac{\partial\rho\psi^{-}\left(w\right)}{w-z_{I}}e^{\phi}\left(z_{I}\right)\sideset{}{^{\prime}}\prod_{r}e^{-\frac{iQ^{2}}{\alpha_{r}}\mathcal{X}^{+}}\left(\hat{\tilde{\mathbf{z}}}_{I^{\left(r\right)}},\hat{\tilde{\bar{\mathbf{z}}}}_{I^{\left(r\right)}}\right)\,.\label{eq:bX}
\end{eqnarray}
Here the prime in $\sideset{}{^{\prime}}\prod$ means that the product
is taken over those $r$ which satisfy 
\begin{equation}
z_{I^{\left(r\right)}}=z_{I}\,,
\end{equation}
and
\begin{eqnarray}
A\left(w\right) & = & -i\partial X^{+}\partial\rho\gamma\left(w\right)-2\partial\left(\partial\rho c\right)\psi^{-}\left(w\right)
\nonumber \\
 &  & -\frac{d-10}{4}i\left[\left(\frac{5\left(\partial^{2}X^{+}\right)^{2}}{4\left(\partial X^{+}\right)^{3}}-\frac{\partial^{3}X^{+}}{2\left(\partial X^{;}\right)^{2}}\right)\left(-2\partial\rho\gamma\right)
-\frac{2\partial^{2}X^{+}}{\left(\partial X^{+}\right)^{2}}\partial\left(-2\partial\rho\gamma\right)
\right. \nonumber \\
& & \hphantom{-\frac{d-10}{4}i\quad}
\left. {}+\frac{\partial^{2}\left(-2\partial\rho\gamma\right)}{\partial X^{+}}-\frac{\left(-2\partial\rho\gamma\right)\partial\psi^{+}\partial^{2}\psi^{+}}{2\left(\partial X^{+}\right)^{3}}\right]\left(w\right)\,.
\end{eqnarray}
Substituting (\ref{eq:bX}) into (\ref{eq:oddBRST3}) and using the
commutators
\begin{eqnarray}
&& \left\{ \hat{Q}^{\prime},\hat{Q}^{\prime}\right\}   =  0\,,\nonumber \\
&& \left\{ \hat{Q}^{\prime},c\left(w_{0}\right)\right\}  =  0\,, \nonumber \\
&& \left\{ \hat{Q}^{\prime},b\left(z\right)\right\}  =  0\,, \nonumber \\
&& \left[\hat{Q}^{\prime},V_{r}^{\left(p,-1\right)}\right] =  0
\qquad  \left(p=-2,-1,0\right)\,,\nonumber \\
&& \left\{ \hat{Q}^{\prime},\oint_{z_{I}}\frac{dz}{2\pi i}\frac{b}{\partial\rho}\left(z\right)e^{\phi}T_{\mathrm{F}}^{\mathrm{LC}}\left(z_{I}\right)\sideset{}{^{\prime}}\prod_{r}e^{-\frac{iQ^{2}}{\alpha_{r}}\mathcal{X}^{+}}\left(\hat{\tilde{\mathbf{z}}}_{I^{\left(r\right)}},\hat{\tilde{\bar{\mathbf{z}}}}_{I^{\left(r\right)}}\right)\right\}   =  0\,, \nonumber \\
&& \left\{ \hat{Q}^{\prime},\oint_{z_{I}}\frac{dz}{2\pi i}\frac{b}{\partial\rho}\left(z\right)\oint_{z_{I}}\frac{dw}{2\pi i}\frac{\partial\rho\psi^{-}\left(w\right)}{w-z_{I}}e^{\phi}\left(z_{I}\right)\sideset{}{^{\prime}}\prod_{r}e^{-\frac{iQ^{2}}{\alpha_{r}}\mathcal{X}^{+}}\left(\hat{\tilde{\mathbf{z}}}_{I^{\left(r\right)}},\hat{\tilde{\bar{\mathbf{z}}}}_{I^{\left(r\right)}}\right)\right\}   =  0\,,~~~~
\end{eqnarray}
we can show that (\ref{eq:oddBRST3}) is equal to 
\begin{eqnarray}
 &  & \int\left[dX^{\mu}d\psi^{\mu}d\bar{\psi}^{\mu}dbd\bar{b}dcd\bar{c}d\beta d\bar{\beta}d\gamma d\bar{\gamma}\right]_{g_{z\bar{z}}^{\mathrm{A}}}e^{-S^{\mathrm{tot}}}\nonumber \\
 &  & \qquad\times\partial\rho c\left(w_{0}\right)\bar{\partial}\bar{\rho}\bar{c}\left(\bar{w}_{0}\right)\nonumber \\
 &  & \qquad\times\prod_{j=1}^{g}\left[\left(\oint_{\alpha_{j}}\frac{dz}{\partial\rho}b_{zz}+\oint_{\alpha_{j}}\frac{d\bar{z}}{\bar{\partial}\bar{\rho}}b_{\bar{z}\bar{z}}\right)\left(\oint_{\beta_{j}}\frac{dz}{\partial\rho}b_{zz}+\oint_{\beta_{j}}\frac{d\bar{z}}{\bar{\partial}\bar{\rho}}b_{\bar{z}\bar{z}}\right)\right]\nonumber \\
 &  & \qquad\times\prod_{I}\oint_{z_{I}}\frac{dz}{2\pi i}\frac{b}{\partial\rho}\left(z\right)\left[-e^{\phi}T_{\mathrm{F}}^{\mathrm{LC}}\left(z_{I}\right)+\frac{1}{4}\oint_{z_{I}}\frac{dw}{2\pi i}\frac{\partial\rho\psi^{-}\left(w\right)}{w-z_{I}}e^{\phi}\left(z_{I}\right)\right]\nonumber \\
 &  & \qquad\times\prod_{I}\left[\oint_{\bar{z}_{I}}\frac{d\bar{z}}{2\pi i}\frac{\bar{b}}{\partial\bar{\rho}}\left(\bar{z}\right)\bar{X}\left(\bar{z}_{I}\right)\right]\prod_{r}e^{-\frac{iQ^{2}}{\alpha_{r}}\mathcal{X}^{+}}\left(\hat{\tilde{\mathbf{z}}}_{I^{\left(r\right)}},\hat{\tilde{\bar{\mathbf{z}}}}_{I^{\left(r\right)}}\right)\nonumber \\
 &  & \qquad\times V_{1}^{\left(-2,-1\right)}\left(Z_{1},\bar{Z}_{1}\right)V_{2}^{\left(0,-1\right)}\left(Z_{2},\bar{Z}_{2}\right)\prod_{r=3}^{N}\left[V_{r}^{\left(-1,-1\right)}(Z_{r},\bar{Z}_{r})\right]\,.\label{eq:oddBRST4}
\end{eqnarray}
The correlation functions of $\psi^{\pm}$ which appear in (\ref{eq:oddBRST4})
can be calculated using (\ref{eq:fermioniccorrodd}) treating $S_{\mathrm{int}}$
perturbatively. 
Since $\partial\rho\left(z_{I}\right)=0$, 
$\oint_{z_{I}}\frac{dw}{2\pi i}
  \frac{\partial\rho\psi^{-}\left(w\right)}{w-z_{I}}$
vanishes unless $\psi^{-}\left(w\right)$ becomes singular at $w=z_{I}$.
Hence if the $\psi^{+}\left(Z_{1}\right)$ in 
$V_{1}^{\left(-2,-1\right)}\left(Z_{1},\bar{Z}_{1}\right)$
is contracted with the $\psi^{-}\left(w\right)$ in 
$\oint_{z_{I}}\frac{dw}{2\pi i}
    \frac{\partial\rho\psi^{-}\left(w\right)}{w-z_{I}}$,
the contour integral over $w$ vanishes. 
Therefore only the terms which involve contraction of the 
$\psi^{+}\left(Z_{1}\right)$ in
$V_{1}^{\left(-2,-1\right)}\left(Z_{1},\bar{Z}_{1}\right)$ with the
$\psi^{-}\left(Z_{2}\right)$ in 
$V_{2}^{\left(0,-1\right)}\left(Z_{2},\bar{Z}_{2}\right)$
survive. 
The $\psi^{-}\left(w\right)$ in 
$\oint_{z_{I}}\frac{dw}{2\pi i}
   \frac{\partial\rho\psi^{-}\left(w\right)}{w-z_{I}}$
should be contracted with $\partial^{n}\psi^{+}\left(z_{I}\right)$
involved in 
$e^{-\frac{iQ^{2}}{\alpha_{r}} \mathcal{X}^{+}}
        \left(\hat{\tilde{\mathbf{z}}}_{I^{\left(r\right)}},
             \hat{\tilde{\bar{\mathbf{z}}}}_{I^{\left(r\right)}}\right)$
or 
$S_{\mathrm{int}}$ but doing so induces another contraction
\begin{equation}
\oint_{z_{J}}\frac{dw}{2\pi i}\frac{1}{w-z_{J}}\partial\rho\begC1{\psi^{-}}\conC{\left(w\right)\partial^{n}}\endC1{\psi^{+}}\left(z_{I}\right)\,,
\end{equation}
with $z_{J}\ne z_{I}$, because 
$e^{-\frac{iQ^{2}}{\alpha_{r}}\mathcal{X}^{+}}
  \left( \hat{\tilde{\mathbf{z}}}_{I^{\left(r\right)}},
         \hat{\tilde{\bar{\mathbf{z}}}}_{I^{\left(r\right)}}\right)$
and $S_{\mathrm{int}}$ are Grassmann even. 
Hence we conclude that
$\oint_{z_{I}}\frac{dw}{2\pi i}
    \frac{\partial\rho\psi^{-}\left(w\right)}{w-z_{I}}$
in (\ref{eq:oddBRST4}) does not contribute to the path integral.
We can do the same thing for the antiholomorphic part and prove that
the $X\left(z_{I}\right),\bar{X}\left(\bar{z}_{I} \right)$ which appear
in (\ref{eq:oddBRST3}) can be replaced by $-e^{\phi}T_{\mathrm{F}}^{\mathrm{LC}}\left(z_{I}\right),-e^{\bar{\phi}}\bar{T}_{\mathrm{F}}^{\mathrm{LC}}\left(\bar{z}_{I}\right)$
for all $I$. By deforming the contours of the antighost insertions
back, we can see that (\ref{eq:ncBRSTodd}) is equal to (\ref{eq:ncBRST2odd}).

\providecommand{\href}[2]{#2}\begingroup\raggedright\endgroup


\begin{thebibliography}{10}

\bibitem{Sen2016c}
A.~Sen, ``{BV Master Action for Heterotic and Type II String Field Theories},''
  \href{http://dx.doi.org/10.1007/JHEP02(2016)087}{{\em JHEP} {\bf 02} (2016)
  087},
\href{http://arxiv.org/abs/1508.05387}{{\tt arXiv:1508.05387 [hep-th]}}.

\bibitem{Sen2015b}
A.~Sen, ``{Gauge Invariant 1PI Effective Action for Superstring Field
  Theory},'' \href{http://dx.doi.org/10.1007/JHEP06(2015)022}{{\em JHEP} {\bf
  1506} (2015)  022},
\href{http://arxiv.org/abs/1411.7478}{{\tt arXiv:1411.7478 [hep-th]}}.

\bibitem{Sen2015j}
A.~Sen, ``{Gauge Invariant 1PI Effective Superstring Field Theory: Inclusion of
  the Ramond Sector},'' \href{http://dx.doi.org/10.1007/JHEP08(2015)025}{{\em
  JHEP} {\bf 08} (2015)  025},
\href{http://arxiv.org/abs/1501.00988}{{\tt arXiv:1501.00988 [hep-th]}}.

\bibitem{Lacroix2017b}
C.~de~Lacroix, H.~Erbin, S.~P. Kashyap, A.~Sen, and M.~Verma, ``{Closed
  Superstring Field Theory and its Applications},''
\href{http://arxiv.org/abs/1703.06410}{{\tt arXiv:1703.06410 [hep-th]}}.

\bibitem{Sen2017}
A.~Sen, ``{Background Independence of Closed Superstring Field Theory},''
\href{http://arxiv.org/abs/1711.08468}{{\tt arXiv:1711.08468 [hep-th]}}.

\bibitem{Zwiebach1993}
B.~Zwiebach, ``{Closed string field theory: Quantum action and the B-V master
  equation},'' \href{http://dx.doi.org/10.1016/0550-3213(93)90388-6}{{\em Nucl.
  Phys.} {\bf B390} (1993)  33--152},
\href{http://arxiv.org/abs/hep-th/9206084}{{\tt arXiv:hep-th/9206084
  [hep-th]}}.

\bibitem{Aoki:1990yn}
K.~Aoki, E.~D'Hoker, and D.~H. Phong, ``{UNITARITY OF CLOSED SUPERSTRING
  PERTURBATION THEORY},''
\href{http://dx.doi.org/10.1016/0550-3213(90)90575-X}{{\em Nucl. Phys.} {\bf
  B342} (1990)  149--230}.

\bibitem{Greensite:1986gv}
J.~Greensite and F.~R. Klinkhamer, ``{NEW INTERACTIONS FOR SUPERSTRINGS},''
\href{http://dx.doi.org/10.1016/0550-3213(87)90256-2}{{\em Nucl. Phys.} {\bf
  B281} (1987)  269}.

\bibitem{Greensite:1987sm}
J.~Greensite and F.~R. Klinkhamer, ``{CONTACT INTERACTIONS IN CLOSED
  SUPERSTRING FIELD THEORY},''
\href{http://dx.doi.org/10.1016/0550-3213(87)90485-8}{{\em Nucl. Phys.} {\bf
  B291} (1987)  557}.

\bibitem{Greensite:1987hm}
J.~Greensite and F.~R. Klinkhamer, ``{SUPERSTRING AMPLITUDES AND CONTACT
  INTERACTIONS},''
\href{http://dx.doi.org/10.1016/0550-3213(88)90622-0}{{\em Nucl. Phys.} {\bf
  B304} (1988)  108}.

\bibitem{Green:1987qu}
M.~B. Green and N.~Seiberg, ``{CONTACT INTERACTIONS IN SUPERSTRING THEORY},''
\href{http://dx.doi.org/10.1016/0550-3213(88)90549-4}{{\em Nucl. Phys.} {\bf
  B299} (1988)  559}.

\bibitem{Wendt:1987zh}
C.~Wendt, ``{SCATTERING AMPLITUDES AND CONTACT INTERACTIONS IN WITTEN'S
  SUPERSTRING FIELD THEORY},''
\href{http://dx.doi.org/10.1016/0550-3213(89)90118-1}{{\em Nucl. Phys.} {\bf
  B314} (1989)  209}.

\bibitem{Ishibashi2017b}
N.~Ishibashi and K.~Murakami, ``{Multiloop Amplitudes of Light-cone Gauge NSR
  String Field Theory in Noncritical Dimensions},''
  \href{http://dx.doi.org/10.1007/JHEP01(2017)034}{{\em JHEP} {\bf 01} (2017)
  034},
\href{http://arxiv.org/abs/1611.06340}{{\tt arXiv:1611.06340 [hep-th]}}.

\bibitem{Ishibashi2017d}
N.~Ishibashi, ``{Light-cone gauge superstring field theory in linear dilaton
  background},'' \href{http://dx.doi.org/10.1093/ptep/ptx012}{{\em PTEP} {\bf
  2017} (2017) no.~3, 033B01},
\href{http://arxiv.org/abs/1605.04666}{{\tt arXiv:1605.04666 [hep-th]}}.

\bibitem{Baba:2009kr}
Y.~Baba, N.~Ishibashi, and K.~Murakami, ``{Light-Cone Gauge Superstring Field
  Theory and Dimensional Regularization},''
  \href{http://dx.doi.org/10.1088/1126-6708/2009/10/035}{{\em JHEP} {\bf 10}
  (2009)  035},
\href{http://arxiv.org/abs/0906.3577}{{\tt arXiv:0906.3577 [hep-th]}}.

\bibitem{Ishibashi:2010nq}
N.~Ishibashi and K.~Murakami, ``{Light-cone Gauge NSR Strings in Noncritical
  Dimensions II -- Ramond Sector},''
  \href{http://dx.doi.org/10.1007/JHEP01(2011)008}{{\em JHEP} {\bf 01} (2011)
  008},
\href{http://arxiv.org/abs/1011.0112}{{\tt arXiv:1011.0112 [hep-th]}}.

\bibitem{D'Hoker:1987pr}
E.~D'Hoker and S.~B. Giddings, ``{UNITARY OF THE CLOSED BOSONIC POLYAKOV
  STRING},''
\href{http://dx.doi.org/10.1016/0550-3213(87)90466-4}{{\em Nucl. Phys.} {\bf
  B291} (1987)  90}.

\bibitem{Giddings:1986rf}
S.~B. Giddings and S.~A. Wolpert, ``{A TRIANGULATION OF MODULI SPACE FROM LIGHT
  CONE STRING THEORY},''
\href{http://dx.doi.org/10.1007/BF01215219}{{\em Commun. Math. Phys.} {\bf 109}
  (1987)  177}.

\bibitem{D'Hoker:1988ta}
E.~D'Hoker and D.~H. Phong, ``{The Geometry of String Perturbation Theory},''
\href{http://dx.doi.org/10.1103/RevModPhys.60.917}{{\em Rev. Mod. Phys.} {\bf
  60} (1988)  917}.

\bibitem{Ishibashi:2013nma}
N.~Ishibashi and K.~Murakami, ``{Multiloop Amplitudes of Light-cone Gauge
  Bosonic String Field Theory in Noncritical Dimensions},''
  \href{http://dx.doi.org/10.1007/JHEP09(2013)053}{{\em JHEP} {\bf 09} (2013)
  053},
\href{http://arxiv.org/abs/1307.6001}{{\tt arXiv:1307.6001 [hep-th]}}.

\bibitem{arakelov}
S.~Arakelov, ``{Intersection Theory of Divisors on an Arithmetic Surface},''
  {\em Math. USSR Izv. 8 1167} (1974)  .

\bibitem{Verlinde1987b}
E.~P. Verlinde and H.~L. Verlinde, ``{Multiloop Calculations in Covariant
  Superstring Theory},''
\href{http://dx.doi.org/10.1016/0370-2693(87)91148-8}{{\em Phys. Lett.} {\bf
  B192} (1987)  95}.

\bibitem{Sen2015}
A.~Sen and E.~Witten, ``{Filling the gaps with PCO's},''
  \href{http://dx.doi.org/10.1007/JHEP09(2015)004}{{\em JHEP} {\bf 09} (2015)
  004},
\href{http://arxiv.org/abs/1504.00609}{{\tt arXiv:1504.00609 [hep-th]}}.

\bibitem{Sen2015a}
A.~Sen, ``{Off-shell Amplitudes in Superstring Theory},''
  \href{http://dx.doi.org/10.1002/prop.201500002}{{\em Fortsch.Phys.} {\bf 63}
  (2015)  149--188},
\href{http://arxiv.org/abs/1408.0571}{{\tt arXiv:1408.0571 [hep-th]}}.

\bibitem{Witten2013a}
E.~Witten, ``{More On Superstring Perturbation Theory: An Overview Of
  Superstring Perturbation Theory Via Super Riemann Surfaces},''
\href{http://arxiv.org/abs/1304.2832}{{\tt arXiv:1304.2832 [hep-th]}}.

\bibitem{Sen2015d}
A.~Sen, ``{Supersymmetry Restoration in Superstring Perturbation Theory},''
  \href{http://dx.doi.org/10.1007/JHEP12(2015)075}{{\em JHEP} {\bf 12} (2015)
  075},
\href{http://arxiv.org/abs/1508.02481}{{\tt arXiv:1508.02481 [hep-th]}}.

\bibitem{Atick1987b}
J.~J. Atick and A.~Sen, ``{Spin Field Correlators on an Arbitrary Genus Riemann
  Surface and Nonrenormalization Theorems in String Theories},''
\href{http://dx.doi.org/10.1016/0370-2693(87)90304-2}{{\em Phys. Lett.} {\bf
  B186} (1987)  339}.

\bibitem{Berkovits:1985ji}
N.~Berkovits, ``{CALCULATION OF SCATTERING AMPLITUDES FOR THE NEVEU-SCHWARZ
  MODEL USING SUPERSHEET FUNCTIONAL INTEGRATION},''
\href{http://dx.doi.org/10.1016/0550-3213(86)90070-2}{{\em Nucl. Phys.} {\bf
  B276} (1986)  650}.

\bibitem{Berkovits:1987gp}
N.~Berkovits, ``{SUPERSHEET FUNCTIONAL INTEGRATION AND THE INTERACTING
  NEVEU-SCHWARZ STRING},''
\href{http://dx.doi.org/10.1016/0550-3213(88)90642-6}{{\em Nucl. Phys.} {\bf
  B304} (1988)  537}.

\bibitem{Baba:2009zm}
Y.~Baba, N.~Ishibashi, and K.~Murakami, ``{Light-cone Gauge Superstring Field
  Theory and Dimensional Regularization II},''
  \href{http://dx.doi.org/10.1007/JHEP08(2010)102}{{\em JHEP} {\bf 08} (2010)
  102},
\href{http://arxiv.org/abs/0912.4811}{{\tt arXiv:0912.4811 [hep-th]}}.

\bibitem{Verlinde:1986kw}
E.~P. Verlinde and H.~L. Verlinde, ``{Chiral bosonization, determinants and the
  string partition function},''
\href{http://dx.doi.org/10.1016/0550-3213(87)90219-7}{{\em Nucl. Phys.} {\bf
  B288} (1987)  357}.

\bibitem{Fay1973}
J.~D. Fay, {\em Theta Functions on Riemann Surfaces}.
\newblock Lecture Notes in Mathematics 352. Springer-Verlag, 1973.

\end{thebibliography}

\end{document}